\documentclass[a4paper,12pt]{article}
\usepackage{amsfonts,amsmath,amssymb,epic,eepic,color,graphicx}

\newfont{\twelvemsb}{msbm10 scaled\magstep1}
\newfont{\eightmsb}{msbm8}
\newfam\msbfam
\textfont\msbfam=\twelvemsb \scriptfont\msbfam=\eightmsb
\catcode`\@=11
\def\Bbb{\ifmmode\let\next\Bbb@\else
\def\next{\errmessage{Use \string\Bbb\space only in math mode}}\fi\next}
\def\Bbb@#1{{\fam\msbfam{{#1}}}}

\newcommand{\be}{\begin{equation}}
\newcommand{\ee}{\end{equation}}
\newcommand{\ba}{\begin{eqnarray}}
\newcommand{\ea}{\end{eqnarray}}

\makeatletter
\renewcommand{\@makecaption}[2]{
   \vskip\abovecaptionskip
   \sbox\@tempboxa{#1. #2}%
   \ifdim \wd\@tempboxa >\hsize
     \small {\bf #1.} #2\par
   \else
     \global \@minipagefalse
     \hb@xt@\hsize{\hfil\box\@tempboxa\hfil}%
   \fi
   \vskip\belowcaptionskip}
\makeatother

\begin{document}
\sloppy
\renewcommand{\thefootnote}{\fnsymbol{footnote}}

\newpage
\setcounter{page}{1}
\begin{flushright}
LAPTH - 1167/06\\
\end{flushright}
\vspace*{0.2 cm} {\it On the 75-th Anniversary of Bethe Ansatz}
\vspace*{1.2 cm}
\begin{center}
{\large \bf Hubbard's Adventures in ${\cal N}=4$ SYM-land?} \\
{\bf Some non-perturbative considerations on finite length operators.}\\
\vspace{1.8cm} {\large G.\ Feverati $^a$\footnote{Financially
supported by INFN.}, D.\ Fioravanti $^b$, P. \ Grinza
$^c$\footnote {Since 1st November at the Department of Theoretical
Physics, University of Torino.} and M.\ Rossi $^{a,b}$
\footnote{E-mails : feverati@lapp.in2p3.fr, fioravanti@bo.infn.it,grinza@to.infn.it,rossi@lapp.in2p3.fr.}}\\
\vspace{.5cm} $^a${\em LAPTH \footnote{UMR 5108 du CNRS,
associ\'ee \`a l'Universit\'e de Savoie.}, 9 Chemin de Bellevue,
BP 110, F-74941 Annecy-le-Vieux Cedex, France} \\
\vspace{.3cm} $^b${\em INFN and Dept. of Physics, University of Bologna, Via Irnerio 46, Bologna, Italy} \\
\vspace{.3cm} $^c${\em LPTA, Universit\'e Montpellier II, Place
Eug\`ene Bataillon, 34095 Montpellier, France}
\end{center}
\renewcommand{\thefootnote}{\arabic{footnote}}
\setcounter{footnote}{0}
\begin{abstract}
{\noindent As} the Hubbard energy at half filling is believed to
reproduce at strong coupling (part of) the all loop expansion of
the dimensions in the $SU(2)$ sector of the planar $ {\cal N}=4$
SYM, we compute an exact non-perturbative expression for it. For
this aim, we use the effective and well-known idea in 2D
statistical field theory to convert the Bethe Ansatz equations
into two coupled non-linear integral equations (NLIEs). We focus
our attention on the highest anomalous dimension for fixed bare
dimension or length, $L$, analysing the many advantages of this
method for extracting exact behaviours varying the length and the
't Hooft coupling, $\lambda$. For instance, we will show that the
large $L$ (asymptotic) expansion is exactly reproduced by its
analogue in the BDS Bethe Ansatz, though the exact expression
clearly differs from the BDS one (by non-analytic terms).
Performing the limits on $L$ and $\lambda$ in different orders is
also under strict control. Eventually, the precision of numerical
integration of the NLIEs is as much impressive as in other
easier-looking theories.
\end{abstract}
\vspace{1cm} {\noindent {\it Keywords}}: Quantum Integrability
(Bethe Ansatz); Non-Linear Integral Equation; Hubbard model; Super
Yang-Mills theories.

\newpage

\section{Prologue}
\setcounter{equation}{0}

It is a modern achievement that gauge theories and in particular
supersymmetric gauge theories {\it hide} many realisations of the
algebraic geometry theorisation (cf. \cite{KW} just as a recent
monumental reference on the {\it last} discovered parallel and
many other features).

More in specific, the AdS/CFT correspondence \cite{MWGKP} should
be a general dictionary, which would equate -- among other
physical objects -- energies of string states to anomalous
dimensions of local gauge-invariant operators of a dual conformal
quantum field theory. Proving or even testing this duality in full
generality may be a formidable task, but the integrability
properties of the ${\cal N}=4$ super Yang-Mills (SYM) theory have
proved to be extremely useful to understand how it may work and to
which extent.

The identification \cite{MZ} of the one-loop dilatation operator
of scalar gauge-invariant fields with bare dimension $L$ with an
$SO(6)$ integrable chain with $L$ sites, reducing in the $SU(2)$
subspace to the spin $1/2$-XXX Heisenberg chain, allowed, by using
the Bethe Ansatz technique \cite{Bethe}, to test the one-loop
AdS/CFT duality in many cases \cite{HLPS,BTZ,GK} beyond the BMN
conditions \cite{BMN}. In the aforementioned cases, special
emphasis has been delivered to the $1/L$ correction, as this would
result as the first quantum correction in string theory; and more
generally all the finite size $L$ corrections would have a similar
stringy origin and importance. Soon afterwards, integrability of
${\cal N}=4$ SYM at higher loops started to be hinted and hunted
\cite {BKS}. After various attempts and tests (cf. for instance
\cite{SS}), eventually in \cite{BDS} an all loop asymptotic
expression has been proposed for the eigenvalues of the dilatation
operator in the $SU(2)$ sub-sector, in terms of the solutions of
Bethe Ansatz-like equations, derived by assuming BMN scaling and
perturbative integrability. Moreover, this proposal (named after
them BDS equations) was shown to give the correct (truncated)
Bethe equations for the five loop dilatation operator, after
deriving the latter as an operator. Nevertheless, the BDS
equations are valid only asymptotically, that is for fixed $L$
when the 't Hooft coupling $\lambda$ is small enough that the
${\cal O} (\lambda^L)$ term ($L$ loops) may be neglected along
with the higher order powers. In fact, the higher loops are
clearly affected by the chain wrapping problem -- namely an
interaction range longer than the chain length -- which is not
taken into account by the BDS proposal. In this respect, an
important progress was the remark, by Rej, Serban and Staudacher
\cite {RSS}, that the $SU(2)$ dilatation operator could be
reproduced up to three loops by the strong coupling expansion of
the Hamiltonian of the half-filled Hubbard model. Many tests of
the proposal \cite{RSS} started to be performed (cf. for instance
\cite{MIN}), while it seems now clear that starting from four
loops the Hubbard model will reproduce only part of the entire
contributions (likely the 'rational ones'), the string
theory/gauge theory discrepancies motivating the introduction of a
specific dressing factor \cite{BES} also in gauge theory. Indeed,
although the dressing factor would also care for the large
$\lambda$ behaviour, it is unclear for now how to insert it into
the two Lieb-Wu Bethe equations for the Hubbard model \cite{LW}.
On the contrary, it is manifest its introduction into the BDS
Bethe Ansatz, and therefore a comparative study of BDS versus
Hubbard model is one of the motivations of this paper.

In a previous paper \cite {FFGR} we proposed a description of the
highest and immediately lower energy states, for both the $SO(6)$
chain and the BDS model, based on the non-linear integral equation
(NLIE). The NLIE was first introduced in \cite {KP, DDV, FMQR} for
studying the finite size scaling of the ground state and of the
excited states in (critical and off-critical) statistical
(lattice) field theories respectively. Although it is equivalent
to the set of all the Bethe Ansatz equations, it is often more
suitable for numerical and analytical calculations, especially
when it is important, like in the present case, to detect how the
anomalous dimension (energy) behaves with the length $L$
(especially in large $L$ investigations). In fact, it condensates
into a single (or only very few) integral equation(s) many
algebraic equations. In this respect, we will prove here that it
is a right tool to deal with the two possible ordering of the
limits of large size $L$ and large coupling $\lambda$.
Furthermore, we will find a systematic way to perform the two
expansions for small coupling and large coupling at any fixed
size.

In this paper we want to introduce the NLIEs as a profitable
treatment of the Hubbard model, especially for the understanding
of the exact scaling behaviour of the dimension (energy) with
$\lambda$ and $L$. We will concentrate on the highest energy
(anomalous dimension) state of the half-filled Hubbard model
($SU(2)$ sub-sector of ${\cal N}=4$ SYM). This state is described
by two coupled NLIEs which will be written in Section 3. In
Section 4 we will give an exact expression for its energy, as a
function of the coupling $g$ and the length of the chain $L$, in
terms of the solution of the two coupled NLIEs. This peculiar
expression for the energy allows a comparison at large length $L$
(Section 5) with the analogous result coming from the BDS chain:
we will show that the $1/L$ leading term and all the next finite
size corrections (power-like and logarithmic) in fact coincide
(i.e. the usual large $L$ asymptotic expansion do coincide), the
difference being captured by exponentially small corrections
(whose leading contribution we estimated at strong coupling).
Moreover, as a consistency check of our findings, in Section 6 the
weak and strong coupling limits of the NLIEs on one hand and of
the energy on the other hand will be studied and shown to
reproduce the known results. The strong coupling for the BDS model
is also carefully analysed. As a consequence, Section 7 is devoted
to the understanding of the ordering of the two distinct limits
$\lambda\rightarrow +\infty$ and $L\rightarrow +\infty$, both in
the Hubbard and BDS models. Eventually, a detailed numerical
analysis is carried out in the last Section 8.

\section{The Hubbard model: a bird's-eye view}

The Hubbard model was introduced as a simplified model for
strongly correlated electrons on a lattice \cite{hub}. In one
dimension, it describes $N_{\text{e}}$ electrons moving on a chain
with $L$ sites and interacting via the Hamiltonian
\begin{equation}
H=-t \sum _{i=1}^L \sum _{\sigma=\uparrow , \downarrow}\left
(c_{i,\sigma}^{\dagger}
c_{i+1,\sigma}+c_{i+1,\sigma}^{\dagger}c_{i,\sigma}\right) +U \sum
_{i=1}^L
c_{i,\uparrow}^{\dagger}c_{i,\uparrow}c_{i,\downarrow}^{\dagger}c_{i,\downarrow}
\, , \label {Hubbham}
\end{equation}
where $c_{i,\sigma}^{\dagger}$, $c_{i,\sigma}$ are (fermionic)
canonical creation and annihilation operators respectively, $t$ is
the strength of the kinetic nearest-neighbour hopping term, $U$
the coupling constant of the density potential and, for our
interests, periodic boundary conditions are assumed, i.e.
$c_{i+L,\sigma}=c_{i,\sigma}$,
$c_{i+L,\sigma}^{\dagger}=c_{i,\sigma}^{\dagger}$.

In the relevant paper \cite {RSS}, a precise sub-set of the
energies $E$ of (\ref{Hubbham}) was conjectured to be proportional
to the anomalous contribution $\gamma$ to the conformal dimensions
in the $SU(2)$ scalar sector of ${\cal N}=4$ super Yang-Mills
theory in the planar limit,
\begin{equation}
\gamma = \frac {\lambda }{8 \pi ^2} E \, , \label{andim}
\end{equation}
provided we restrict ourselves to the half-filling case
$N_{\text{e}}=L$ and also equate the length $L$ to the number of
constituent operators.

Now, a very important part of the correspondence between an
integrable system and a gauge theory is the mapping of the
coupling constants, and the latter can be easily argued to be a
strong-weak coupling duality for many reasons\footnote{A simple
one is the limiting case of the Hubbard model in strong coupling
(and half filling) \cite{BEM}, i.e. the $1/2-$XXX revealed,
originally, at one loop by Minahan and Zarembo \cite{MZ}.}.
Therefore, one possible choice, reproducing the known results up
to three loops, may well be \cite {RSS}
\begin{equation}
t=-\frac {1}{{\sqrt {2}}g}=-\frac {2\pi}{\sqrt {\lambda}} \, ,
\quad U=-\frac1{g^2}= -\frac {8\pi^2}{\lambda} \, , \label{tUg}
\end{equation}
where $\lambda=Ng_{YM}^2=8\pi ^2 g^2$ is the 't Hooft coupling of
the $SU(N)$ SYM theory in the planar limit ($N\rightarrow
\infty$). But this can be modified by higher order contributions
still preserving, of course, the matching outcomes. Actually, to
have a loop expansion of (\ref{andim}) in $g^2$ with the right
{\it wrapping phenomenon} occurring at ${\cal O}(g^{2L})$, we need
to introduce a constant magnetic flux $\phi$ \cite{RSS},
\begin{equation}
H=\frac {1}{{\sqrt {2}}g} \sum _{i=1}^L \sum _{\sigma=\uparrow ,
\downarrow}\left (e^{i\phi }c_{i,\sigma}^{\dagger}
c_{i+1,\sigma}+e^{-i\phi}c_{i+1,\sigma}^{\dagger}c_{i,\sigma}\right)
-\frac {1}{g^2} \sum_{i=1}^L
c_{i,\uparrow}^{\dagger}c_{i,\uparrow}c_{i,\downarrow}^{\dagger}c_{i,\downarrow}
\, , \label {Hubbham2}
\end{equation}
distinguishing odd and even lengths: $\phi=0$ when $L$ is odd and
$\phi=\frac {\pi}{2L}$ when $L$ is even.

The Hubbard hamiltonian (\ref{Hubbham}) describes an integrable
model (infinite many conserved charges in involution), which was
diagonalised by Lieb and Wu by Bethe Ansatz in 1968 \cite{LW}. The
twisted Hamiltonian (\ref{Hubbham2}) is still integrable and the
Lieb-Wu equations easily generalise. In the half-filling case they
read \cite {RSS}
\begin{eqnarray}
e^{i\hat{k}_jL}&=&\prod _{l=1}^M \frac {u_l - \frac {2t}{U}\sin
  (\hat{k}_j+\phi)-\frac {i}{2}} {u_l - \frac {2t}{U}\sin
  (\hat{k}_j+\phi)+\frac {i}{2}} \nonumber \\
\prod _{j=1}^L \frac {u_l - \frac {2t}{U}\sin
  (\hat{k}_j+\phi)+\frac {i}{2}} {u_l - \frac {2t}{U}\sin
  (\hat{k}_j+\phi)-\frac {i}{2}}&=& \mathop{\prod _ {m=1}}_{m \not=l}^M \frac
  {u_l-u_m+i}{u_l-u_m-i} \, , \label {Beqs}
\end{eqnarray}
where $M$ is the number of down spins. The spectrum of the
Hamiltonian is then given in terms of the pseudo-momenta
$\hat{k}_j$, by the {\it free dispersion relation}
\begin{equation}\label{energia}
E=-2 t \sum _{j=1}^{L} \cos (\hat{k}_j+\phi) \,.
\end{equation}
Starting from here, we will equivalently derive two coupled
nonlinear integral equations (NLIEs) for the antiferromagnetic
state of the model at any value of $L$, thanks to the methods used
in \cite{FFGR}. For reason of completeness, we point out that the
thermodynamics (infinite length $L$, but finite temperature) of
the Hubbard model has been studied \cite {KB,JKS} by means of
three NLIEs (for a summary of the procedure and a complete list of
references see \cite{DEGKKK}). This approach was based on the
equivalence of the (quantum) one-dimensional Hubbard model with
the (classical) two-dimensional Shastry model. For the gauge
theory understanding, we need to obtain energies of the Hubbard
model at zero temperature, but at any value of the length $L$.
This completely justifies our approach, and {\it a fortiori} in
the perspective of extending our calculations to excited states
(i.e. lower dimension operators in the SYM spectrum).

\section{Two non-linear integral equations (NLIEs)}
\setcounter{equation}{0}

Looking at the Bethe equations (\ref {Beqs}), we define the
function
\begin{equation}
\Phi (x,\xi)=i \ln \frac {i\xi +x}{i\xi -x} \, , \label {Phi}
\end{equation}
with the branch cut of $\ln(z)$ along the real negative $z$-axis
in such a way that $-\pi < \arg z <\pi$. Then, we perform a gauge
transformation which amounts to adding the magnetic flux:
\begin{equation}
k_j=\hat{k}_j+\phi \,.
\end{equation}
After a possible choice of the counting functions as
\begin{eqnarray}
W(k)&=&L(k-\phi) -\sum _{l=1}^M \Phi \left (u_l-\frac {2t}{U}\sin
k,
\frac{1}{2} \right ) \, , \label {Wdef} \\
Z(u)&=&\sum _{j=1}^L \Phi  \left (u - \frac {2t}{U}\sin k_j,
\frac{1}{2} \right ) -\sum _{m=1}^M \Phi \left (u-u_m, 1 \right )
\, , \label {Zdef}
\end{eqnarray}
we can rewrite the Bethe equations, by taking their logarithm, in
the usual form of quantisation conditions for the Bethe roots
$\{k_j,u_l\}$,
\begin{eqnarray}
W(k_j)&=&\pi M +2 \pi I^w_j \, , \\
Z(u_l)&=&\pi (M-L+1+2I^z_l) \, .
\end{eqnarray}
From now on, we specialise our treatment to the highest energy
state, consisting of the maximum number $M=L/2$ of real roots
$u_l$ and of $L$ real roots $k_j$. For simplicity reasons, we
restrict ourselves to the case $M\in 2{\Bbb N}$ (the remaining
case $M\in 2{\Bbb N}+1$ is a simple modification of this case),
which obviously implies $L\in 4{\Bbb N}$.

In the definition of the counting functions (\ref{Wdef},
\ref{Zdef}) we have to deal with sums of functions computed on
real Bethe roots, $k_j$ and $u_l$. Let us first concentrate on
functions of $k_j$. We notice that $k_j$ may run only within the
first Brillouin zone $[-\pi,\pi)$ and that the functions of $k_j$
involved are periodic with period $2\pi$. On the other hand, the
counting function $W(k)$ is quasi-periodic on that interval and
$e^{iW(k)}$ and $W'(k)$ are indeed periodic. Using the Cauchy
theorem to circulate the interval $[-\pi,\pi)$ by a small
displacement $\epsilon>0$ (this periodic case has been developed
in \cite{FR}), we get
\begin{eqnarray}\label{cauchyW}
\sum _{j=1}^L f(k_j)=&-&\int _{\pi }^{-\pi} \frac{dk}{2\pi
i}f(k+i\epsilon) \frac
{iW^{\prime}(k+i\epsilon)e^{iW(k+i\epsilon)}}{1-e^{iW(k+i\epsilon)}}-
\nonumber \\
&-&\int _{-\pi}^{\pi} \frac{dk}{2\pi i}f(k-i\epsilon) \frac
{iW^{\prime}(k-i\epsilon)e^{iW(k-i\epsilon)}}{1-e^{iW(k-i\epsilon)}}
\, ,
\end{eqnarray}
where the two complex integrals along $-\epsilon<{\mbox { Im}}
k<\epsilon$ at ${\mbox {Re}} k=\pm\pi$ have been neglected thanks
to the periodicity properties of $f(k)$ and $e^{iW(k)}$. A thumb
rule to understand this {\it logarithmic indicator formula} goes
as follows: since $W^\prime (k)> 0$ \footnote{We have numerical
evidence for that; in Section 5, we analytically prove this
statement when $L\rightarrow \infty$.}, the first integral is
simply the logarithmic derivative in the following formula, but
the second one is not because of the non-analyticity of the
logarithm \footnote {We use the approximation
$e^{iW(k+i\epsilon)}=e^{iW(k)}e^{-\epsilon W'(k)}$: therefore, we
suppose $\epsilon \ll 1$.}. Nevertheless, the latter can be simply
manipulated into a logarithmic derivative of an analytic function
plus an extra piece:
\begin{eqnarray}
\sum _{j=1}^L f(k_j)&=&-\int _{-\pi}^{\pi} \frac {dk}{2\pi
i}f(k+i\epsilon)
\frac {d}{dk}\ln \left [1-e^{iW(k+i\epsilon)}\right]+ \\
&+& \int _{-\pi}^{\pi} \frac {dk}{2\pi i}f(k-i\epsilon) \frac
{d}{dk}\ln \left [1-e^{-iW(k-i\epsilon)}\right] + \int
_{-\pi}^{\pi} \frac {dk}{2\pi} f(k-i\epsilon) W^\prime
(k-i\epsilon)  \, . \nonumber
\end{eqnarray}
To make the last term useful, we can compute it along the real
axis without any harm, because of the periodicity of $f(k)$ and
$W'(k)$; then, after integrating by parts the two integrals before
it, we arrive at
\begin{eqnarray}
\sum _{j=1}^L f(k_j)&=&\int _{-\pi}^{\pi} \frac {dk}{2\pi
i}f^\prime
(k+i\epsilon) \ln \left [1-e^{iW(k+i\epsilon)}\right]- \label {Wsum0}\\
&-&\int _{-\pi}^{\pi} \frac {dk}{2\pi i}f^\prime (k-i\epsilon) \ln
\left [1-e^{-iW(k-i\epsilon)}\right] +\int _{-\pi}^{\pi} \frac
{dk}{2\pi} f(k)W^\prime (k) \, , \nonumber
\end{eqnarray}
because the boundary terms vanish as a consequence of the
periodicity of $f(k)$ and $e^{iW(k)}$. Upon integrating by parts
the last term, we finally obtain
\begin{eqnarray}
\sum _{j=1}^L f(k_j)&=&-\int _{-\pi}^{\pi} \frac {dk}{2\pi}
f^\prime (k)W(k) + {\mbox { Im}} \int _{-\pi}^{\pi} \frac {dk}{\pi
  }f^\prime (k+i\epsilon ) \ln \left [
  1-e^{iW(k+i\epsilon )}\right] + \nonumber \\
&+& \left [ \frac {f(k)W(k)}{2\pi} \right ]^{\pi}_{-\pi} \, .
\label {Wsumeps}
\end{eqnarray}
We will mainly use such formula in the $\epsilon \rightarrow 0^+$
limit,
\begin{equation}
\sum _{j=1}^L f(k_j)=-\int _{-\pi}^{\pi} \frac {dk}{2\pi} f^\prime
(k)W(k) + \int _{-\pi}^{\pi} \frac {dk}{\pi
  }f^\prime (k) {\mbox { Im}} \ln \left [
  1-e^{iW(k+i0)}\right] + \left [ \frac {f(k)W(k)}{2\pi} \right
]^{\pi}_{-\pi} \, , \label {Wsum}
\end{equation}
which reads as (\ref {Wsum}) because of the supposed analyticity
of $f(k)$ on the real axis.

For what concerns a sum of a generic function $g(x)$ for $x$ being
any root $u_l$ (which is in principle everywhere in the real axis
for the ground state\footnote{As it is well-known after \cite{DDV,
FMQR}, for other states complex roots and holes have to be
included.}), we can go along similar steps and repeat the original
procedure for $x\in \mathbb{R}$ \cite{DDV, FMQR}. In this case the
boundary terms appearing during the computations can be neglected
thanks to different applicable reasons. One sufficient set of
conditions, which apply to the case $g=\Phi$, relevant for the
derivations of the NLIEs for $W$ and $Z$, turns out to be
\footnote {Obviously, $\epsilon$ appearing in the equations for
$Z$ is different from the homonymous constant related to $W$.}
\begin{eqnarray}
&&Z'(\pm \infty + iy)=0 \, , \, -\epsilon <y<\epsilon \, ; \quad g(+\infty)=-g(-\infty) \, , \quad Z(+\infty)=-Z(-\infty) \, , \nonumber \\
&& \quad Z(+\infty\pm i\epsilon)=-Z(-\infty\mp i\epsilon ) \, ,
\quad g(+\infty\pm i\epsilon)=-g(-\infty\mp i\epsilon ) . \label
{fZcond}
\end{eqnarray}
In formul\ae \, we can write
\begin{equation}
\sum _{l=1}^M g(u_l)=-\int _{-\infty}^{\infty} \frac
     {dx}{2\pi}g^{\prime}(x)Z(x)+{\mbox { Im}} \int _{-\infty}^{\infty}\frac
     {dx}{\pi}g^{\prime}(x+i\epsilon ) \ln \left
[1+e^{iZ(x+i\epsilon )}\right ] \, , \label {Zsumeps}
\end{equation}
or, in the $\epsilon \rightarrow 0^+$ limit,
\begin{equation}
\sum _{l=1}^M g(u_l)=-\int _{-\infty}^{\infty} \frac
     {dx}{2\pi}g^{\prime}(x)Z(x)+\int _{-\infty}^{\infty}\frac
     {dx}{\pi}g^{\prime}(x){\mbox { Im}} \ln \left
[1+e^{iZ(x+i0 )}\right ] \, . \label {Zsum}
\end{equation}
Eventually, we have our building blocks in formul\ae \,
(\ref{Wsum}) and (\ref{Zsum}), where the r.h.s. is written through
each counting function respectively. Let us apply them to the
definition of the $W(k)$,
\begin{eqnarray}
W(k)&=&L(k-\phi)+\int _{-\infty}^{\infty} \frac
     {dx}{2\pi}\Phi ^ {\prime} \left ( x-\frac
{2t}{U} \sin k, \frac {1}{2} \right ) Z(x) - \nonumber \\
&-& \int _{-\infty}^{\infty}\frac
     {dx}{\pi} \Phi ^{\prime} \left ( x-\frac
{2t}{U} \sin k, \frac {1}{2} \right ) {\mbox {Im}}\ln \left
[1+e^{iZ(x+i0)}\right ] \, , \label {Weq1}
\end{eqnarray}
and to the definition of $Z(u)$,
\begin{eqnarray}
Z(u)&=&L \Phi \left (u,\frac {1}{2}\right )+ \frac {2t}{U}\int
_{-\pi}^{\pi} \frac {dk}{2\pi}\Phi ^{\prime}
\left( u-\frac {2t}{U}\sin k, \frac {1}{2} \right) \cos k \: W(k)-\nonumber \\
&-&\frac {2t}{U}\int _{-\pi}^{\pi} \frac {dk}{\pi}\Phi
^{\prime}\left ( u-\frac {2t}{U}\sin k, \frac {1}{2} \right ) \cos
k
{\mbox { Im}}\ln \left [1-e^{iW(k+i0)}\right ]- \label {Zeq1} \\
&-& \int _{-\infty}^{\infty}\frac
     {dy}{2\pi} \Phi ^{\prime} (u-y,1)Z(y)
+ \int _{-\infty}^{\infty}\frac
     {dy}{\pi} \Phi ^{\prime} (u-y,1){\mbox { Im}}\ln \left
[1+e^{iZ(y+i0)}\right ] \, . \nonumber
\end{eqnarray}
Inserting in the equation for $Z$ the expression for $W$ coming
from (\ref {Weq1}), we get
\begin{eqnarray}
Z(u)&=&L \Phi \left (u ,\frac {1}{2}\right ) + L \frac {2t}{U}\int
_{-\pi}^{\pi} \frac {dk}{2\pi}\Phi ^{\prime}\left (
u-\frac {2t}{U}\sin  k , \frac {1}{2} \right ) \cos k \:(k-\phi) -\nonumber \\
&-& \int _{-\infty}^{\infty}\frac
     {dy}{2\pi} \Phi ^{\prime} (u-y,1)\, Z(y)
+ \int _{-\infty}^{\infty}\frac
     {dy}{\pi} \Phi ^{\prime} (u-y,1){\mbox { Im}}\ln \left
[1+e^{iZ(y+i0)}\right ] -  \nonumber   \\
&-&\frac {2t}{U}\int _{-\pi}^{\pi} \frac {dk}{\pi}\Phi
^{\prime}\left ( u-\frac {2t}{U}\sin k , \frac {1}{2} \right )
\cos k  {\mbox { Im}} \ln \left [1-e^{iW(k+i0)}\right ] \, ,
\label {Zeq2}
\end{eqnarray}
where we used the following cancellation of terms,
\begin{equation}
\int _{-\pi}^{\pi} dk~ \Phi ^{\prime}\left ( x-\frac {2t}{U}\sin k
, \frac {1}{2} \right ) \cos k ~ \Phi ^{\prime}\left ( y-\frac
{2t}{U}\sin k , \frac {1}{2} \right ) =0 \, ,
\end{equation}
which can be easily proven by performing the change of variable
$k\rightarrow \pi - k$. We now write the equation for $Z(u)$ (\ref
{Zeq2}) in terms of Fourier transforms\footnote{We define the
Fourier transform $\hat{f}(p)$ of a function $f(x)$ as given by
\begin{equation}\hat{f}(p) = \int_{-\infty}^{\infty} dx ~ e^{-ipx} f(x)\, . \end{equation}}, using
\begin{equation}
\hat \Phi (p, \xi)=\frac {2\pi} {i}P\left (\frac {1}{p}\right)
e^{-\xi |p|} \, ,
\end{equation}
where $P$ indicates the principal value distribution. We obtain
the following expression,
\begin{eqnarray}
\hat Z(p)&=& L 2 \pi \frac{e^{-\frac {|p|}{2}} }{i}P\left (\frac
{1}{p}\right) + L \frac {2t}{U} \int _{-\pi}^{\pi} dk\  e^{-i\frac
{2 t p}{U}\sin k }  e^{-\frac {|p|}{2}}
\cos k  ~(k-\phi) - \nonumber \\
&-& e^{-|p|} \hat Z(p) +2 e^{-|p|} \hat L _Z (p) -\frac {4t}{U}
e^{-\frac {|p|}{2}} \int _{-\pi}^{\pi} dk\
e^{-i\frac {2tp}{U}\sin k } \cos k ~ L_W(k) = \nonumber \\
&=& L \frac {2\pi}{i}P\left (\frac {1}{p}\right) e^{-\frac
{|p|}{2}} J_0\left (\frac{2tp}{U}\right )
- e^{-|p|} \hat Z(p)+ 2 e^{-|p|} \hat L _Z (p)- \nonumber \\
&-&\frac {4t}{U}  e^{-\frac {|p|}{2}} \int _{-\pi}^{\pi} dk \,
e^{-i\frac {2tp}{U}\sin k } \cos k~ L_W(k) \, ,
\end{eqnarray}
where we used the integral definition of the Bessel function
$J_0(z)$,
\begin{equation}
J_0(z)=\int _{-\pi}^{\pi} \frac {dk}{2\pi}\, e^{i\,z\sin k}\,,
\end{equation}
and also the following shorthand notations
\begin{equation}
L_W(k)= {\mbox {Im}}\ln \left [1-e^{iW(k+i0)}\right ] \, , \quad
L_Z(x)= {\mbox {Im}}\ln \left [1+e^{iZ(x+i0)}\right ] \, .
\end{equation}
The terms proportional to $\hat Z(p) $ are now collected and
reorganized as
$$
\hat Z(p)= L \frac {\pi}{i}P\left (\frac {1}{p}\right) \frac
{J_0\left (\frac
    {2tp}{U}\right )}{\cosh \frac {p}{2}} +
\frac {2}{1+e^{|p|}}\hat L _Z (p) -\frac {2t}{U}  \frac {1}{\cosh
\frac {p}{2}} \int _{-\pi}^{\pi} dk~ e^{-i\frac {2tp}{U}\sin k}
\cos k \: L_W(k) \, ,
$$
and, coming back to the 'coordinate' space, we obtain the first of
two nonlinear integral equations for our counting functions,
\begin{eqnarray}
Z(u)&=&L \int _{-\infty}^{\infty} \frac {dp}{2p} \sin (pu) \frac
{J_0\left ( \frac {2tp}{U}\right )}{\cosh \frac {p}{2}}+2 \int
 _{-\infty}^{\infty} dy \ G(u-y) \ {\mbox {Im}}\ln \left
[1+e^{iZ(y+i0)}\right ]- \nonumber \\
&-&\frac {2t}{U}\int _{-\pi}^{\pi} dk \cos k \frac {1}{\cosh \left
( \pi u - \frac {2t\pi}{U}\sin  k \right ) } \
 {\mbox {Im}}\ln \left [1-e^{iW(k+i0)}\right ] \, , \label {Zeq4}
\end{eqnarray}
where $G(x)$ is the same kernel function that appears in the spin
$1/2$-XXX chain and in the BDS Bethe Ansatz (eq. 2.24 of
\cite{FFGR}),
\begin{equation}
G(x)=\int _{-\infty}^{\infty} \frac {dp}{2\pi} e^{ipx} \frac
{1}{1+e^{|p|}} \, . \label {Gxxx}
\end{equation}
We notice that the first line of the NLIE for $Z$ (\ref {Zeq4})
coincides with the NLIE (eq. 3.15 of \cite{FFGR}) for the counting
function of the highest energy state of the BDS model. The second
line of (\ref {Zeq4}) is the genuine contribution of the Hubbard
model.

We finally remark that NLIE (\ref {Zeq4}) can be written in the
alternative form
\begin{eqnarray}
Z(u)&=&L \int _{-\pi}^{\pi}\frac {dk}{2\pi} {\mbox { gd}}\left (
\pi u - \frac {2t\pi}{U} \sin k \right )+2 \int
 _{-\infty}^{\infty} dy \ G(u-y) \ {\mbox {Im}}\ln \left
[1+e^{iZ(y+i0)}\right ]+ \nonumber \\
&+&  \int _{-\pi}^{\pi}\frac {dk}{\pi}\ \frac {d}{dk} {\mbox {gd}}
\left ( \pi u - \frac {2t\pi}{U} \sin k \right ){\mbox {Im}}\ln
\left [1-e^{iW(k+i0)}\right ] \, ,
\end{eqnarray}
after introducing the hyperbolic amplitude (the Gudermannian)
${\mbox {gd}}(x)$:
\begin{equation}
{\mbox {gd}}(x)=\int _{0}^{x} \frac {dt}{\cosh t}=2\arctan e^x -
\frac {\pi}{2} \, .
\end{equation}
On the other hand, starting from (\ref {Weq1}) and inserting in it
the equation for $Z$ (\ref {Zeq4}), we obtain the second of our
nonlinear integral equations:
\begin{eqnarray}
W(k)&=&L\left[ (k-\phi) + \int _{-\infty}^{\infty} \frac {dp}{p}
\sin \left ({\frac
  {2tp}{U}\sin k }\right ) \frac {J_0\left (\frac {2tp}{U}\right )}{1+e^{|p|}}\right]
- \nonumber \\
&-&\int _{-\infty}^{\infty}dx \, \frac {1}{\cosh \left ( \frac
{2t\pi}{U}\sin k-\pi x \right ) } \,
{\mbox {Im}}\ln \left[1+e^{iZ(x+i0)}\right ]-  \label {Weq2}\\
&-& \frac {4t}{U} \int _{-\pi}^{\pi} dh \ G \left ( \frac {2t}{U}
\sin h-\frac {2t}{U}\sin k \right ) \cos h \mbox{ Im} \ln
\left[1-e^{iW(h+i0)}\right ] \, . \nonumber
\end{eqnarray}
The two equations (\ref{Zeq4}, \ref{Weq2}) are coupled by integral
terms and are completely equivalent to the Bethe equations for the
highest energy state.

\section{The energy or anomalous dimension.}
\setcounter{equation}{0}

The eigenvalues of the Hamiltonian (\ref {Hubbham2}) on the Bethe
states are given by (\ref{energia}). The highest eigenvalue can be
worked out by using (\ref {Wsum}). We get:
\begin{equation}
E=-2t \left\{ \int _{-\pi}^{\pi} \frac {dk}{2\pi}\sin k ~W(k)
-\int _{-\pi}^{\pi} \frac {dk}{\pi}\sin k {\mbox { Im}}\ln
\left[1-e^{iW(k+i0)}\right ] - L \right\} \,. \label {En1}
\end{equation}
We now insert the NLIE for $W$ (\ref{Weq2}) and observe the
cancellation of the first and last terms:
\begin{equation}
\int _{-\pi}^{\pi} \frac {dk}{2\pi} \sin k ~(k-\phi) -1=0 \, .
\end{equation}
Therefore, we are left with
\begin{eqnarray}
E&=&-2t \left\{ L \int _{-\pi}^{\pi} \frac {dk}{2\pi}\sin k  \int
_{-\infty}^{\infty} \frac {dp}{p}\sin \left (\frac {2tp}{U} \sin k
\right) \frac {J_0\left ( \frac {2tp}{U} \right)}{e^{|p|}+1}
\right.
\nonumber \\
&-& \int _{-\pi}^{\pi} \frac {dk}{2\pi}\sin k  \int
_{-\infty}^{\infty}dx \frac {L_Z(x)}{\cosh \left (\frac {2 \pi
t}{U}
  \sin k -\pi x \right )}  \nonumber \\
&-&\frac{2t}{U}  \int _{-\pi}^{\pi} \frac {dk}{\pi}\sin k  \int
_{-\pi}^{\pi} dh~ G \left [ \frac {2t}{U} (\sin h-\sin k)
\right ]\cos h  ~L_W(h)   \nonumber \\
&-&\left . \int _{-\pi}^{\pi} \frac {dk}{\pi}\sin k ~ L_W(k)
\right\} \, .
\end{eqnarray}
We recognize the presence of the Bessel function
\begin{equation}
J_1(z)=\frac {1}{2\pi i}\int _{-\pi}^{\pi} dk \sin k ~ e^{iz
  \sin k} \, , \label {J1}
\end{equation}
in the first three terms of the right hand side (in second and
third we have used the Fourier representations, e.g. (\ref {Gxxx})
for $G$). We finally obtain that the highest eigenvalue of (\ref
{Hubbham2}) is expressed in terms of the counting functions $Z$
and $W$ as follows,
\begin{eqnarray}
E&=&-2t\left\{ L \int _{-\infty}^{\infty} \frac {dp}{p} \frac
{J_0\left (\frac
  {2tp}{U}\right ) J_1\left (\frac  {2tp}{U}\right ) }{e^{|p|}+1}
+  \int _{-\infty}^{\infty} dx \left [ \int _{-\infty}^{\infty}
\frac
  {dp}{2\pi}\frac {e^{ipx}}{\cosh \frac {p}{2}}i J_1\left (\frac
  {2tp}{U}\right ) \right ]  L_Z(x) - \right. \nonumber  \\
&-& \frac{2t}{U} \left. \int _{-\pi}^{\pi} \frac {dh}{\pi} L_W(h)
\cos h \left [ \int _{-\infty}^{\infty} \frac {dp}{i}
 e^{i\frac {2tp}{U}\sin h } \frac {J_1\left (\frac
  {2tp}{U}\right ) }{e^{|p|}+1} \right ]-
\int _{-\pi}^{\pi} \frac {dh}{\pi} L_W(h)
\sin h  \right\} \nonumber \\
&\equiv& E_L+E_Z+E_{W1}+E_{W2} \,, \quad \text{and}\quad E_W\equiv
E_{W1}+E_{W2}.  \label {Eexp}
\end{eqnarray}
The first line of (\ref {Eexp}), namely $E_L+E_Z$, coincides
formally with the expression of the highest energy of the BDS
chain as given in equation (3.24) of \cite{FFGR}. However, we have
to remember that for the Hubbard model $Z$ satisfies a NLIE which
is different from that of the BDS model. On the other hand, the
second line, i.e. $E_W=E_{W1}+E_{W2}$, is a completely new
contribution.

\section{The large $L$ expansions of Hubbard and BDS energies in comparison \label{comparison}}
\setcounter{equation}{0}

As it was first noticed by \cite {RSS}, in the $L=\infty$ limit
(thermodynamic limit) the leading term of the highest energy
$E_{\text{BDS}}$ of the BDS model coincides with the
thermodynamical expression of the Hubbard model, the first
contribution in (\ref {Eexp}). Since we can provide exact
expressions for energies at any length $L$, we want to extract
more information about the difference $E_{\text{BDS}}-E$ when $L$
is large, but finite. And we are in the position to obtain this
for any value of the coupling constant $g$.

For the highest energy $E_{\text{BDS}}$, many detailed results
were given in \cite{FFGR}, where it was expressed as (cf. equation
3.24)
\begin{eqnarray}
E_{\text{BDS}}&=&\frac{\sqrt{2}}{g} \left\{ L \int
_{-\infty}^{\infty} \frac {dp}{p} \frac { J_0 ( \sqrt{2}g p ) J_1
(\sqrt{2}g p ) }{e^{|p|}+1} +  \right. \nonumber \\
&+& \left. \int _{-\infty}^{\infty} dx \left [  \int
_{-\infty}^{\infty} \frac
  {dp}{2\pi}\frac {e^{ipx}}{\cosh \frac {p}{2}}i J_1 (\sqrt{2}g p) \right]
 L_{Z_{\text{BDS}}}(x) \right\} \label {EBDS} \, ,
\end{eqnarray}
(with the usual shorthand $L_{Z_{\text{BDS}}}(x)= {\mbox {Im}}\ln
[1+e^{iZ_{\text{BDS}}(x+i0)} ]$), in terms of the solution of the
NLIE
\begin{equation}
Z_{\text{BDS}}(x)=L \int _{-\infty}^{\infty} \frac {dp}{2p} \sin
px \frac {J_0 ( \sqrt{2}g p )}{\cosh \frac {p}{2}}+2 \int
 _{-\infty}^{\infty} dy \ G(x-y) \ {\mbox {Im}}\ln \left
[1+e^{iZ_{\text{BDS}}(y+i0)}\right ].
\end{equation}
We use in this section the parametrization (\ref {tUg}) and we
focus our attention on the energy formula (\ref{Eexp}). For the
purposes of this section, it is convenient to restore a finite
(but small) value for the parameter $\epsilon>0$, used in the
treatment of the function $W$.

As an effect of that, the last term of (\ref {Zeq4}) becomes
\begin{equation}
-\sqrt{2}g \ {\mbox {Im}} \int _{-\pi}^{\pi} dk \cos (k+i\epsilon)
\frac {1}{\cosh \left [ \pi u - \pi \sqrt{2}g \sin (k+i\epsilon)
\right ] } \
 \ln \left [1-e^{iW(k+i\epsilon)}\right ] \, . \label{Zeq4last}
\end{equation}
On the other hand, the last term of the NLIE (\ref {Weq2}) for $W$
takes the form
\begin{equation}
- 2\sqrt{2}g \mbox{ Im} \int _{-\pi}^{\pi} dh \ G \left [
\sqrt{2}g \sin (h+i\epsilon)-\sqrt{2}g \sin k \right ] \cos
(h+i\epsilon) \ln \left[1-e^{iW(h+i\epsilon)}\right ] \, .
\end{equation}
Finally, $E_{W_1}$ and $E_{W_2}$ (\ref {Eexp}) are rewritten as
\begin{eqnarray}
&& E_{W_1}=-2\mbox{ Im} \int _{-\pi}^{\pi} \frac {dh}{\pi}   \cos
(h+i\epsilon) \left [ \int _{-\infty}^{\infty} \frac {dp}{i}
 e^{i\sqrt{2}g p \sin (h+i\epsilon) }  \frac {J_1\left (
  \sqrt{2}g p \right ) }{e^{|p|}+1} \right ] \ln \left[1-e^{iW(h+i\epsilon)}\right ]  \nonumber  \\
&& E_{W_2}=-\frac {\sqrt{2}}{g} \mbox{ Im} \int _{-\pi}^{\pi}
\frac {dh}{\pi} \sin (h+i\epsilon)  \ln
\left[1-e^{iW(h+i\epsilon)}\right ] \, . \nonumber
\end{eqnarray}
All these formul{\ae} depend on $L$ through the function $\ln
\left[1-e^{iW(k+i\epsilon)}\right ]$. Therefore, we have to study
such a function when $L$ is large. Since $\epsilon \ll 1$ (see
Footnote 2), we can approximate, at first order,
\begin{equation}\label{appr}
\ln [1-e^{iW(k+i\epsilon)}] = \ln [1-e^{iW(k)-\epsilon
W^{\prime}(k)}]+ {\cal O}(\epsilon ^2) \, .
\end{equation}
If we suppose also that
\begin{equation}\label{approx}
\epsilon\, W'(k) \gg 1  \qquad \forall ~ k\in [-\pi, \pi] \, ,
\end{equation}
(this condition will be better stated in few lines), then the
factor $\exp [-\epsilon \, W'(k)]$ becomes very small and we are
led to the final approximation:
\begin{equation}
\ln [1-e^{iW(k+i\epsilon)}] \simeq - e^{iW(k)} e^{-\epsilon W'(k)}
\, .
\end{equation}
On the other hand, when $L\rightarrow \infty$ we can approximate
$W(k)$ by its 'forcing term',
\begin{equation}
W(k)\simeq L\left[ k+ \int _{-\infty}^{\infty} \frac {dp}{p} \sin
\left (
  \sqrt{2}g p \sin k \right )\frac {J_0\left (\sqrt{2}g p \right )}{1+e^{|p|}} \right]\, ,
\end{equation}
and, consequently, its derivative by
\begin{equation}
W'(k)\simeq L \left[ 1+ {\sqrt{2}} g\cos k \int
_{-\infty}^{\infty} dp ~  \cos \left(\sqrt{2}g p \sin k\right)
\frac {J_0 (\sqrt{2}g p )}{1+e^{ |p|}} \right]\, .
\end{equation}
The function in the square brackets has a minimum at $k=\pm \pi$,
which we call $\omega (g)$:
\begin{equation}
\omega (g)= 1- {\sqrt{2}} g \int _{-\infty}^{\infty} dp ~  \frac
{J_0 (\sqrt{2}g p )}{1+e^{ |p|}} = 1-2 \int _{0}^{\infty} dx \frac
{J_0 (x)}{1+e^{\frac {x}{\sqrt {2}g} }}
 \, .
\end{equation}
Expanding the denominator in power series and integrating term by
term we get
\begin{equation}
\omega (g)= 1- 2 \int _{0}^{\infty} dx J_0 (x)\sum _{n=1}^{\infty}
(-1)^{n+1} e^{-n\frac {x}{\sqrt {2}g} } = 1- 2 \sum
_{n=1}^{\infty} (-1)^{n+1} \frac {1}{\sqrt {1+\frac {n^2}{2g^2}}}
\, .
\end{equation}
The last expression can be seen as a result of an integration in
the complex plane
\begin{equation}
\omega (g)= 1 + \int _{\Gamma }\frac {dz}{i}  \frac {1}{\sin \pi
z} \frac {1}{\sqrt {1+\frac {z^2}{2g^2}}} \, ,
\end{equation}
on a curve $\Gamma $ (see Figure 6.4 of Takahashi's book \cite
{TAK}), which surrounds the poles on the positive real axis of
$\frac {1}{\sin \pi z}$, excluding the origin. We can deform the
integration contour to the curve consisting of the points $\delta
+ iy $, with $\delta >0$ fixed and $|y|>\rho>0$, and of a
semicircle of radius $\rho$ around the origin; then we let $\delta
$ and $\rho$ go to zero. The pole at $z=0$ gives a contribution
$-1$ to the integral in the previous formula. The integral on the
points $|y|<{\sqrt {2}g}$ is zero by disparity of the integrand.
On the other hand, the integrand computed for $\bar y>{\sqrt
{2}g}$ equals the integrand in $-\bar y$, because they contain
square roots of complex numbers of the same modulus, but lying
just above (for $\bar y>{\sqrt {2}g}$) or just below (for $\bar
y<-{\sqrt {2}g}$) the cut. Therefore we are left with
\begin{equation}
\omega (g)= 2\int _{{\sqrt {2}g} }^{\infty} dy  \frac {1}{\sinh
\pi y} \frac  {1}{\sqrt {\frac {y^2}{2g^2}-1}} \, .
\end{equation}
From this expression, it easily follows that $\omega (g) >0$.
Moreover, one can show that $\omega (0)=1$ and that
\begin{equation}
g \rightarrow \infty \, \Rightarrow  \omega (g) \simeq g \, {\mbox
{exp}}(-\pi {\sqrt {2}} g) \, .
\end{equation}
We conclude that $\forall \, g $ the derivative $W'$ is everywhere
greater than a positive constant: $W'(k)>L~ \omega (g)>0, \,
\forall ~ k \in [-\pi,\pi]$. As a consequence of this, the assumed
conditions on $\epsilon $ and $L$ can be stated as
\begin{equation}
\frac {1}{L~\omega (g)} \ll \epsilon \ll 1 \, . \label {epsicond}
\end{equation}
We now consider the two $W$-depending terms of (\ref {Eexp}),
$E_{W1}$ and $E_{W2}$, when (\ref {epsicond}) holds. We have the
following inequalities:
\begin{gather}
|E_{W1}| \leq  2  \int _{-\pi}^{\pi} \frac {dh}{\pi}~|\cos
(h+i\epsilon)|~ e^{-\epsilon W'(h)}~\left |
\int_{-\infty}^{\infty} \frac {dp}{i}
 e^{i\sqrt{2}g p \sin (h+i\epsilon)} \frac {J_1\left (\sqrt{2}g p \right ) }{e^{|p|}+1}
\right | \nonumber \\
< 2 e^{-\epsilon L\, \omega (g)}\int _{-\pi}^{\pi} \frac
{dh}{\pi}~|\cos (h+i\epsilon) |~ \left | \int_{-\infty}^{\infty}
\frac {dp}{i} e^{i\sqrt{2}g p \sin (h+i\epsilon)} \frac {J_1\left
(
  \sqrt{2}g p \right ) }{e^{|p|}+1} \right | \, . \label {Emaggio}
\end{gather}
The integral contained in this last line is finite, as far as
$\epsilon$ is sufficiently small: it should be $\sinh \epsilon <
\frac {1}{\sqrt{2}g}$ and this condition is always satisfied, as
we will show in the following Remark 1. Therefore, we conclude
that, in the limit $L\rightarrow \infty$,
\begin{equation}
|E_{W_1}|<  f_1(g)\, e^{-\epsilon L\, \omega (g)} \, , \label
{ew1}
\end{equation}
where we have indicated with $f_1$ the function of $g$ (but not of
$L$) appearing in (\ref {Emaggio}). The same conclusion for
$E_{W2}$,
\begin{equation}
|E_{W_2}|<  f_2(g)\, e^{-\epsilon L\, \omega (g)} \, , \label
{ew2}
\end{equation}
can be obtained, in the limit $L\rightarrow \infty$, by a similar
reasoning.

\medskip

On the other hand, the same procedure can be applied to the third
term of the r.h.s. of the NLIE for $Z$ (\ref {Zeq4}), which we
have rewritten for finite $\epsilon$ in (\ref {Zeq4last}). This
term marks the difference between the NLIE for $Z$ in the Hubbard
model and for $Z_{\text{BDS}}$ in the BDS Ansatz and acts a a
forcing term in the NLIE for the difference $Z_{\text{BDS}}-Z$.
One concludes that, in the limit $L\rightarrow \infty$, such a
term is exponentially small and, consequently, that
\begin{equation}
|Z_{\text{BDS}}(x)-Z(x)|< f_Z(x, g)\, e^{-\epsilon L\, \omega (g)}
\, , \label {zdiff}
\end{equation}
with an analogous meaning of the function $f_Z(x,g)$.

\medskip

Now, we turn to the expression for the highest energy in the
Hubbard model (\ref {Eexp}) and discuss its relation with the
analogous one (\ref {EBDS}) in the BDS context, when $L$ is large.
We remark that the second term in the r.h.s. of (\ref {Eexp}) is
formally identical to the second term of (\ref {EBDS}), the only
difference being that in the latter $Z$ is replaced by
$Z_{\text{BDS}}$. However, the result (\ref {zdiff}) implies that
the difference between these two terms is indeed smaller than
$f_{E_Z}(g)\, e^{-\epsilon L\, \omega (g)}$, with $f_{E_Z}$ a
positive function of $g$. This finding, together with (\ref {ew1},
\ref {ew2}), allows us to state that, for all finite values of
$g$,
\begin{equation}
L\rightarrow \infty \quad \Rightarrow \, |E_{\text{BDS}}-E| <
f_E(g)\, e^{-\epsilon L\, \omega (g)} \, , \label {E-EBDS}
\end{equation}
i.e. the difference between the highest energies in the Hubbard
and in the BDS model is exponentially small at large $L$.
Therefore, not only their leading terms coincide, but also all the
{\em power-like and logarithmic finite size corrections}: this is
exactly the usual asymptotic expansion for large volume in
statistical field theory. As a confirmation of this statement, we
observe that the $1/L$ correction to the highest energy of the BDS
model, found in \cite {FFGR} and expressed in terms of the
modified Bessel functions $I_0,\,I_1$ by
\begin{equation}\label{fsc}
\frac{\sqrt2}{L\pi g}\ \frac{I_1(\sqrt2 \pi g)}{I_0(\sqrt2 \pi
g)}\ \frac{\pi^2}{6} \, ,
\end{equation}
exactly matches the same result for the Hubbard model, obtained
with different methods by \cite{woynar, DEGKKK}. This has been
studied numerically in Fig.~\ref{unosul}.

{\bf Remark 1.} The variable $\epsilon>0$ introduced in
(\ref{cauchyW}) has to satisfy the condition $\epsilon \ll 1$ (see
Footnote 2). In any case, an upper bound for $\epsilon$ comes from
the condition that the integration contour of (\ref {cauchyW})
contains no singularities of the functions $f(x)$ appearing in the
integrand. As far as the NLIE for $W$ is concerned, the function
appearing in the integrations is $\Phi$ (\ref{Wdef}). Therefore,
the singularities come from terms like
\begin{equation}
\log \left( \frac{i}{2} \pm (u_l-\frac{2t}{U} \sin k) \right)\,.
\end{equation}
More precisely, $\bar {k}$ is a singularity if
\begin{equation}
\frac{1}2 \pm [-\frac{2t}{U} \mbox{ Im}(\sin \bar{k})]  =0\,,
\qquad u_l-\frac{2t}{U} \mbox{ Re}(\sin \bar{k}) =0\,.
\end{equation}
We concentrate on the first equation that takes the form
\begin{equation}\label{dominata}
|\sinh (\mbox{Im }\bar{k})| = \frac{U}{4t}~ \frac1{|\cos (\mbox{Re
} \bar{k})|}\geq \frac{U}{4t} \,.
\end{equation}
The upper bound for $\epsilon$, $\epsilon _M>\epsilon $, is given
by the smallest value of $\mbox{Im }\bar{k}$, namely
\begin{equation}\label{largeeps}
\epsilon_{M} = \mbox{arcsinh}{\frac{U}{4t}} = \mbox{arcsinh} \frac
{1}{2 \sqrt{2}g}\,.
\end{equation}

{\bf Remark 2.}  When $g\ll 1$, we already know that
$E_{\text{BDS}}-E = {\cal {O}}(g^{2L})= {\cal {O}}(e^{2L\ln g})$.
In consequence of that, statement (\ref{E-EBDS}) is already known
to be valid when $g \ll 1$. The results of this section allow to
extend the validity of (\ref {E-EBDS}) -- for the highest energy
state -- also to the non-perturbative region.

{\bf Remark 3.} On the other hand, when $g\gg 1$, we can give an
explicit expression for the estimated difference (\ref {E-EBDS}).
More precisely, we can exactly evaluate $E-E_{\text{BDS}}$ in the
double limit $L\rightarrow \infty$, $g\rightarrow \infty$.

When $g\rightarrow \infty$, we have $\epsilon \leq \epsilon _M
=\frac {1}{2{\sqrt {2}}g}\ll 1$. Performing the $g\rightarrow
\infty$ limit of the $L\rightarrow \infty$ limit of
$W(k+i\epsilon)$, we get
\begin{equation}
W(k+i\epsilon)\simeq L [ k+i\epsilon +\arcsin \sin k +i\epsilon \,
{\mbox {sgn}} (\cos k )]\, .
\end{equation}
The choice $\epsilon =\epsilon _M$ allows this expression to be an
expansion in powers of $\frac {1}{g}$, exact up to terms ${\cal
O}(1/g)$. In the same limit, the $p$-integral contained in the
formula for $E_{W_1}$ becomes
\begin{equation}
\int _{-\infty}^{\infty} \frac {dp}{i}
 e^{i\sqrt{2}g p \sin (h+i\epsilon) } \frac {J_1\left (
  \sqrt{2}g p \right ) }{e^{|p|}+1} \rightarrow  \frac {1}{{\sqrt {2}}g} \frac {\sin h }{|\cos h|} \, .
\end{equation}
It follows that in the double limit $L\rightarrow \infty$,
$g\rightarrow \infty$,
\begin{eqnarray}
E_{W_1}&=& -\frac {{\sqrt {2}}}{g} \mbox{ Im} \int _{-\pi}^{\pi}
\frac {dh}{\pi} \sin h \, {\mbox {sgn}} (\cos h )
\, \ln \left [ 1-e^{iL (h+\arcsin \sin h )-\epsilon L (1+ {\text {sgn}} \cos h ) }\right ] \, , \nonumber \\
E_{W_2}&=& -\frac {{\sqrt {2}}}{g} \mbox{ Im} \int _{-\pi}^{\pi}
\frac {dh}{\pi} \sin h \, \ln \left [1- e^{iL (h+\arcsin \sin h
)-\epsilon L (1+ {\text {sgn}} \cos h )} \right ] \, . \nonumber
\end{eqnarray}
Therefore,
\begin{eqnarray}
E_{W_1}+E_{W_2}&=&-2\frac {{\sqrt {2}}}{g} \mbox{ Im} \int
_{-\pi/2}^{\pi/2} \frac {dh}{\pi} \sin h \, \ln \left
[1- e^{2iLh-2\epsilon L} \right ] \simeq \\
&\simeq & 2\frac {{\sqrt {2}}}{g} e^{-2\epsilon L }  \int
_{-\pi/2}^{\pi/2} \frac {dh}{\pi} \sin h \,
\sin 2Lh  \simeq \nonumber \\
&\simeq & - \frac {2{\sqrt {2}}}{\pi L g}  e^{ -\frac {L}{{\sqrt
{2}}g}}  \, , \nonumber
\end{eqnarray}
where we have kept only the leading term proportional to $1/L$ and
we have chosen $\epsilon = \epsilon _M=\frac {1}{2{\sqrt {2}}g}$.

On the other hand, the term (\ref {Zeq4last}), which marks the
difference between $Z$ and $Z_{\text {BDS}}$, in the double limit
$L\rightarrow \infty$, $g\rightarrow \infty$ becomes
\begin{eqnarray}
&& - \mbox{ Im} \int _{-\pi}^{\pi} dk \cos k \, \delta (\sin k) \,
\ln \left [1-e^{iL (k+\arcsin \sin k )-\epsilon
L (1+{\text {sgn}} \cos k )} \right ] = \nonumber \\
&=& - \mbox{ Im}  \left [ \ln \left (1-e^{-2 \epsilon L }\right )-
\ln \left (1-e^{iL\pi} \right ) \right ] =0
\end{eqnarray}
Therefore, in the double limit $L\rightarrow \infty$,
$g\rightarrow \infty$ we have $Z=Z_{\text {BDS}}$ and,
consequently,
\begin{equation}
E-E_{\text {BDS}}=E_{W_1}+E_{W_2}=-\frac {2{\sqrt {2}}}{\pi L g}
e^ {-\frac {L}{{\sqrt {2}}g} }  \, . \label {lglimit}
\end{equation}
This behaviour is typical of the "wrapping effects". A similar
results in the context of string theory was found in \cite {NZZ}.

{\bf Remark 4.} For intermediate values of $g$ we can not make a
prediction for the "velocity" of the exponential damping at large
$L$ of $|E_{\text{BDS}}-E|$. However, numerical data in
Section~\ref{loafg} are consistent with (\ref {E-EBDS}).

\section{Two limiting regimes: strong and weak coupling.}
\setcounter{equation}{0}

Conversely to the previous Section, we want now to explore the
Hubbard energy (\ref{Eexp}) in two limiting regimes, $\frac{t}{U}
\rightarrow 0$ and $\frac{U}{t} \rightarrow 0$ for any fixed $L$.
They define, respectively, the strong and the weak coupling in the
Hubbard model and allow for simplifications and comparison between
our results and analogous ones obtained by other methods. These
computations are also useful as tests for the NLIEs of $W$
(\ref{Weq2}) and of $Z$ (\ref{Zeq4}). Besides, we analyse the
analogous limit $g\rightarrow +\infty$ of the BDS energy for any
fixed value of $L$.

\subsection{Strong coupling limit in the Hubbard model, i.e. weak coupling in SYM: large $\frac{U}{t}$.}

A well known result of the perturbative expansion of the Hubbard
Hamiltonian around $\frac{t}{U}=0$ at (strong) half filling shows
that the leading term is the Heisenberg $1/2$-XXX spin chain
Hamiltonian \cite{A}. This Section is devoted to derive how our
formalism consistently reproduces this result and makes natural a
linear expansion beyond this order. For this aim it is crucial to
observe that the NLIE for $W$ becomes redundant for very small
$\frac{t}{U}=0$. Indeed, the NLIE for $Z(x)$ (\ref{Zeq4}) easily
reduces to
\begin{eqnarray}
Z(x)&=&L \int _{-\infty}^{\infty} \frac {dp}{2p} ~\frac{\sin px}
{\cosh \frac {p}{2}}+2 \int
 _{-\infty}^{\infty} dy\ G(x-y) {\mbox { Im}}\ln \left
[1+e^{iZ(y+i0)}\right ] + {\cal O}\left (\frac {t}{U} \right )
\nonumber \\
&=& L \ {\mbox {gd }} \pi x  +2 \int
 _{-\infty}^{\infty} dy\ G(x-y) {\mbox { Im}}\ln \left
[1+e^{iZ(y+i0)}\right ] + {\cal O}\left (\frac {t}{U} \right )\, ,
\end{eqnarray}
and hence it precisely agree with the single NLIE for the spin
$1/2$-XXX chain (equation (2.25) of \cite{FFGR}) upon forgetting
the ${\cal O}\left (\frac {t}{U} \right )$ terms, namely
\begin{equation}
Z(x)=Z_{\text{XXX}}(x)+ {\cal O}\left (\frac {t}{U} \right ) \, .
\label {Zlimit}
\end{equation}
In the same limit we evaluate the terms entering the rhs of the
NLIE for $W$ (\ref {Weq2}). The integration term on the first line
behaves as follows:
\begin{eqnarray}
&& L  \int _{-\infty}^{\infty} \frac {dp}{p}\ \frac {\sin \left
(\frac
  {2tp}{U}\sin  k \right )}{1+e^{|p|}} J_0 \left (\frac
  {2tp}{U}\right)=\frac {2tL}{U} \sin k\int
  _{-\infty}^{\infty}dp \frac {1}{1+e^{|p|}} + {\cal O}\left (\frac
  {t^3}{U^3} \right ) \nonumber \\
  &=& \frac {4tL}{U} \sin k \ln 2 + {\cal O}\left (\frac
  {t^3}{U^3} \right ) \, .
\end{eqnarray}
The term on the second line contains $L_Z$ and is (at least) $
{\cal O}\left (\frac
  {t}{U}\right)$, since
\begin{equation}
\int _{-\infty}^{\infty} dx ~ \frac {L_{Z_{\text{XXX}}}(x) }
{\cosh  \pi x} =0 \, .
\end{equation}
The two terms just computed are enough to distinguish the leading
order of $W$,
\begin{equation}
W(k)=L(k-\phi)+ {\cal O}\left (\frac  {t}{U} \right ) \, . \label
{leadingW}
\end{equation}
This result can be used in the NLIE (\ref {Zeq4}) for $Z$ to show
that the third term of the r.h.s. is $ {\cal O}\left (\frac
  {t^2}{U^2}\right)$, because
\begin{equation}
\int _{-\pi}^{\pi} dk \cos k {\mbox { Im}} \ln \left
[1-e^{iL(k-\phi+i0 )}\right ]=0 \, .
\end{equation}
Since also the first term of the r.h.s. of (\ref {Zeq4}) is $
{\cal O}\left (\frac
  {t^2}{U^2}\right)$, we can correct (\ref {Zlimit}) as
\begin{equation}
Z(x)=Z_{\text{XXX}}(x)+ {\cal O}\left (\frac {t^2}{U^2} \right )
\, . \label {Zlimit2}
\end{equation}
Consequently, the term on the second line of (\ref {Weq2}) is
\begin{equation}
-  \int _{-\infty}^{\infty} dx  \frac {L_Z(x)}{\cosh \left [ \frac
{2 t\pi}{U}\sin k -\pi x \right ] } = -\frac {2t\pi}{U} \sin k
\int
  _{-\infty}^{\infty} dx ~ \frac {L_{Z_{\text{XXX}}}(x) \sinh \pi x}{\cosh ^2 \pi x}
 + {\cal O}\left (\frac {t^2}{U^2} \right )\, .
\end{equation}
For what concerns the third line of (\ref {Weq2}), we use (\ref
{leadingW}) to get
\begin{gather}
- \frac {4t}{U} \int _{-\pi}^{\pi} dh ~G \left [ \frac {2t}{U}
(\sin h-\sin k ) \right] \cos h  {\mbox { Im}}\ln \left [1-e^{i W(h+i0)}\right ]= \\
= - \frac {4t}{U}\, G(0)\int _{-\pi}^{\pi} dh\ \cos h {\mbox {
Im}} \ln \left [1-e^{iL(h-\phi+i0)}\right ] + {\cal O}\left (\frac
  {t^2}{U^2} \right )=  {\cal O}\left (\frac  {t^2}{U^2} \right ) \, . \nonumber
\end{gather}
The integral in the last line vanishes, as one can see from the
following calculation:
\begin{gather}
\int_{-\pi}^{\pi}dh~ \cos h ~\ln
\frac{1-e^{iL(h-\phi+i0)}}{1-e^{-iL(h-\phi-i0)}}=
\int_{-\pi}^{\pi}dh~ \cos (h+\phi) ~\ln
\frac{1-e^{iL(h+i0)}}{1-e^{-iL(h-i0)}}=
\nonumber \\
=\int_{-\pi}^{\pi}dh~ (\cos h \cos \phi -\sin h \sin \phi) ~\ln
\frac{1-e^{iL(h+i0)}}{1-e^{-iL(h-i0)}}=0-0=0 \,. \label{integrale}
\end{gather}
The summand containing $\cos h \cos \phi$ is odd under the change
$h\rightarrow -h$, so its integral is zero. The remaining term is
also odd under the change $h\rightarrow \pi-h$, thanks to the
parity of $L$. Therefore, we conclude that in the limit
$\frac{t}{U}\rightarrow 0$ the solution of (\ref {Weq2}) becomes
\begin{equation}
W(k)=L(k-\phi) + \frac {4tL}{U} \ \sin k  \ln 2 -\frac
{2t\pi}{U}\sin k \int_{-\infty}^{\infty} dx  \frac
{L_{Z_{\text{XXX}}}(x) \sinh \pi x}{\cosh ^2 \pi x}
 + {\cal O}\left (\frac {t^2}{U^2} \right ) \, . \label {Wlim1}
\end{equation}
Curiously enough, we recognize in (\ref {Wlim1}) the highest
energy $E_{\text{AFX}}$ of the ferromagnetic spin $1/2$-XXX chain
(eq. 2.33 of \cite {FFGR}):
\begin{equation}
W(k)=L(k-\phi) +\frac {2t}{U} \sin k \ E_{\text{AFX}}  + {\cal
O}\left (\frac
  {t^2}{U^2} \right ) \, . \label {Wlim2}
\end{equation}
We now compute the highest energy (\ref {Eexp}) in that limit.
Using (\ref {Zlimit}), it is easy to see that the first line of
(\ref{Eexp}), $E_L+E_Z$, is proportional to the highest energy
$E_{\text{AFX}}$ of the ferromagnetic spin $1/2$-XXX chain:
\begin{equation}\label{energiaxxx}
 E_L+E_Z = -2t \left [ \frac {t}{U} E_{\text{AFX}} + {\cal O}\left (\frac
  {t^3}{U^3} \right ) \right ] \,.
\end{equation}
Among the remaining terms, the first one, $E_{W1}$, is $2t\, {\cal
O}\left (\frac
  {t^3}{U^3} \right )$: therefore, it can be neglected, with respect to $E_L+E_Z$, in the same limit.

The last term to evaluate, $E_{W2}$, deserves more attention. We
evaluate it using (\ref {Wlim2}) and expanding the logarithm where
it appears:
\begin{eqnarray}
E_{W2}&=& 2t \int _{-\pi}^{\pi} \frac {dh}{\pi}\sin h \,
\frac1{2i}
\ln \frac{1-e^{iW(h+i0)}}{1-e^{-iW(h-i0)}}  \nonumber \\
&=& 2t \left [ \int _{-\pi}^{\pi} \frac {dh}{\pi}\sin h \,
\frac1{2i} \ln \frac{ 1-e^{iL(h-\phi+i0)+i \frac {2t}{U} \sin
(h+i0) E_{\text{AFX}}}} { 1-e^{-iL(h-\phi-i0)-i \frac {2t}{U} \sin
(h-i0) E_{\text{AFX}}}}
+{\cal O}\left(\frac{t^2}{U^2}\right) \right ] \nonumber \\
&=&2t \left \{ \int _{-\pi}^{\pi} \frac  {dh}{\pi} \  \sin h \,
\frac {1}{2i}\ln \frac{1-e^{iL(h-\phi+i0)}}{1-e^{-iL(h-\phi-i0)}}+\right. \nonumber \\
&+& \left. \frac {t}{U} E_{\text{AFX}} \int _{-\pi}^{\pi} \frac
{dh}{\pi} \ \sin^2 h \left [ \frac {-e^{iL (h-\phi+i0)}}{1-{e^{iL
(h-\phi+i0)}}} -\frac {e^{-iL (h-\phi-i0)}}{1-{e^{-iL
(h-\phi-i0)}}}\right ] +{\cal O}\left(\frac{t^2}{U^2}\right)
\right \} \, . \nonumber
\end{eqnarray}
Now, we now shift the integration variable $h\rightarrow h+\phi$
\begin{eqnarray}
E_{W2}&=&2t \left [ \int _{-\pi}^{\pi}\frac  {dh}{\pi} \ \sin
\left (h+\phi \right)\frac {1}{2i}\ln \frac
      {1-e^{iL(h+i0)}}{1-e^{-iL(h-i0)}}+ \right.\nonumber \\
&+&  \left. \frac {t}{U} \frac {E_{\text{AFX}}}{iL} \int
_{-\pi}^{\pi} \frac  {dh}{\pi} \  \sin ^2  (h+\phi) \frac
{d}{dh}\ln \frac {1-e^{iL(h+i0)}}{1-e^{-iL(h-i0)}} +  {\cal
O}\left(\frac{t^2}{U^2}\right) \right ] \, .
\end{eqnarray}
The first integral is very similar to (\ref{integrale}) and it
vanishes in the same way as (\ref{integrale}) does. After an
integration by parts, the second integral is brought into the form
\begin{eqnarray}
&&  -2t  \ \frac {t}{U} \frac {E_{\text{AFX}}}{iL} \int
_{-\pi}^{\pi} \frac  {dh}{\pi} \ \sin [2(h+\phi)]\ \ln \frac
{1-e^{iL(h+i0)}}{1-e^{-iL(h-i0)}} = \nonumber \\
&& -2t \ \frac {t}{U} \frac {E_{\text{AFX}}}{iL} \int
_{-\pi}^{\pi} \frac  {dh}{\pi} (\sin 2h \cos 2\phi +\cos 2h \sin
2\phi ) \ln \frac {1-e^{iL(h+i0)}}{1-e^{-iL(h-i0)}} \, . \nonumber
\end{eqnarray}
The second integral is zero by parity; the first one can be shown
to vanish by using the change of variable $h \rightarrow \frac
{\pi}{2} - h$, remembering (Section 2) that $L\in 4{\Bbb N}$. We
conclude that, when $\frac{t}{U} \rightarrow 0$, $E_{W2}$ is $2t
\, {\cal O}\left(\frac{t^2}{U^2}\right)$: so, it can be neglected
with respect to $E_L+E_Z$. We conclude that in the limit $t/U
\rightarrow 0$ the energy (\ref {Eexp}) behaves as
\begin{eqnarray}
E &=&-2t \left \{ \frac {t}{U} \left [ 2L\ln 2  + \int
_{-\infty}^{\infty }dy \left(\frac {d}{dy} \frac {1}{\cosh {\pi
y}}\right) L_{Z_{\text{XXX}}}(y) \right ]
+ {\cal O}\left(\frac{t^2}{U^2}\right) \right \}= \nonumber \\
&=& -2t \left [ \frac {t}{U} E_{\text{AFX}}+  {\cal
O}\left(\frac{t^2}{U^2}\right) \right ] \, , \label{Exxx}
\end{eqnarray}
i.e. the Hubbard energy coincides -- except for an overall factor
-- with the same quantity of the ferromagnetic spin $1/2$-XXX
chain (equation 2.33 of \cite{FFGR}). Besides, with the
parametrization (\ref {tUg}) the overall constant in (\ref{Exxx})
$-2t^2/U=1$. Therefore, in that case we have the exact coincidence
\begin{equation}
\lim _{\frac {t}{U} \rightarrow 0} E=E_{\text{AFX}} \, ,
\end{equation}
which actually encodes all the non-linearity of this expansion,
being the rest, which we omit here, just a linear order by order
addition to this.

\subsection{Weak coupling limit in the Hubbard model, i.e. strong coupling in SYM: small $\frac{U}{t}$.}
\label{strHubb}

On the contrary, we now expand the Hubbard energy around
$\frac{U}{t}=0$ in a systematic way, though we will stop at the
first perturbative order. In fact, by means of the usual (time
independent) perturbation theory the first terms of the expansion
were produced by \cite{MV}. Instead, here we study the NLIE for
$W$ (\ref {Weq2}) by expanding
\begin{equation}
W(k;\frac{U}{t})=W_0(k)+ \frac {U}{t} W_1(k)+
o\left(\frac{U}{t}\right) \, .
\end{equation}
The second term of the r.h.s. of (\ref {Weq2}) is expanded as
\begin{gather}
 L \int _{-\infty}^{+\infty} \frac {dp}{p} \, \sin (
  p\sin k )\frac {J_0\left (
  {p}\right )}{1+e^{\frac {U|p|}{2t}}}=L \int _{-\infty}^{+\infty} \frac {dp}{p} \, \sin (
  p\sin k )\frac {J_0\left (
  {p}\right )}{2}+ o\left(\frac{U}{t}\right) = \nonumber \\
= L \arcsin \sin k + o\left(\frac{U}{t}\right) \, ,
\end{gather}
since the order ${\cal O}(U/t)$ contribution vanishes:
\begin{equation}
-\frac {UL}{4t} \int _{0}^{+\infty} dp \sin (
  p\sin k ) J_0(p) =0 \, .
\end{equation}
The third term on the right hand side is
\begin{eqnarray}
&-&\int _{-\infty}^{+\infty}dx \, \frac {L_Z(x)}{\cosh \left [
\frac {2t\pi}{U}\sin k -\pi x \right ] }   =-\frac{U}{2t}\int
_{-\infty}^{+\infty}dx \left [ \int _{-\infty}^{+\infty} \frac
{dp}{2\pi} \frac {e^{ip\left (\sin k -\frac {U}{2t}x
\right)}}{\cosh \frac {pU}{4t}} \right ] L_Z(x)
= \nonumber \\
&=&- \frac {U}{2t} \delta(\sin k) \int _{-\infty}^{+\infty}dx~
L_{Z_0}(x) + {\cal O}\left(\frac{U^2}{t^2}\right) \, ,
\label{zetazero}
\end{eqnarray}
where $Z_0$ indicates the order zero of $Z$ in the limit
$U/t\rightarrow 0$.

For what concerns the fourth term, we obtain
\begin{eqnarray}
&-& \frac {4t}{U} \int _{-\pi}^{\pi} dh \, G \left [ \frac
{2t}{U}\sin
  h-\frac {2t}{U}\sin
  k \right ] \cos h \, L_W(h) = \nonumber \\
&-& \int _{-\pi}^{+\pi}dh \, \cos h \, \delta (\sin h - \sin k) \,
\left [ L_{W_0}(h) - \frac {U}{t} \, {\mbox {Re}} \, \frac
{e^{iW_0(h+i0)}}{1-e^{iW_0(h+i0)}} \, W_1(h) +
o\left(\frac{U}{t}\right)\right ]  \, , \nonumber
\end{eqnarray}
since as a distribution
\begin{equation}
\lim _{U/t \rightarrow 0} \frac {2t}{U} G\left (\frac {2t}{U} x
\right ) = \frac {1}{2} \delta (x) + {\cal
O}\left(\frac{U^2}{t^2}\right) \, . \label{Gdelta}
\end{equation}

Therefore, at the order zero in $U/t$ the NLIE for $W$ reduces to
\begin{eqnarray}
W_0(k)&=&L(k-\phi) + L \int _{-\infty}^{+\infty} \frac {dp}{2p}
\sin (p\sin k) J_0(p)-\nonumber \\
&-&\int _{-\pi}^{+\pi}dh \cos h \, \delta (\sin h-\sin k) \,
L_{W_0}(h)= \label
{Weqappr} \\
&=& L(k-\phi)+L \, {\mbox {arcsin}} \sin k - {\mbox {sgn}}(\cos k)
\, [ L_{W_0}(k) - L_{W_0}(\pi {\mbox {sgn}}k -k)] \, . \nonumber
\end{eqnarray}
In the last step we used formula (6.693.1) from \cite{GR}.

Now, it is not difficult to show that the solution of
(\ref{Weqappr}) is
\begin{equation}
W_0(k)=L(k-\phi)+2\pi N(k) \, , \label {W_0}
\end{equation}
where the function $N(k)$ takes only integer values. As far as
energy calculations (\ref {Eexp}) are concerned, the knowledge of
$N(k)$ is not required.

\medskip

Let us now focus ourselves on the NLIE for $Z$ (\ref {Zeq4}).
After the change of variable $p\rightarrow pU/2t$ in the integrand
of the forcing term, we can rewrite it as follows:
\begin{eqnarray}
Z(x)&=&L \int _{-\infty}^{+\infty} \frac {dp}{2p} \, \sin \frac
{pUx}{2t}\, \frac{J_0\left (  {p} \right )}{\cosh \frac{pU}{4t}
}+2 \int_{-\infty}^{+\infty} dy \, G(x-y) \, {\mbox {Im}}\ln \left
[1+e^{iZ(y+i0)}\right ]-\nonumber  \\
&-&\frac{2t}{U}\int _{-\pi}^{\pi} dk \, \cos k  \frac{1}{\cosh
\left [ \pi x - \frac{2t\pi}{U}\sin k \right ] }\,
 {\mbox {Im}}\ln \left
[1-e^{iW(k+i0)}\right ] \, . \label {Zeq5}
\end{eqnarray}
When $\frac{U}{t}\rightarrow 0$ the forcing term is clearly ${\cal
O}(U/t)$. In order to estimate the last term we express the
inverse of the $\cosh $ function in terms of its Fourier transform
and we get
\begin{eqnarray}
&-&  \frac{2t}{U}\int _{-\pi}^{\pi} dk \, \cos k  \int _{-\infty}^{\infty}\frac{dp}{2\pi}{e^{ip\left (x-\frac {2t}{U}\sin k \right)}} \frac {1}{{\cosh \frac{p}{2}}} L_W(k)= \nonumber \\
&=&- \int _{-\pi}^{\pi} dk \, \cos k  \int _{-\infty}^{\infty}\frac {dp}{2\pi} \frac {e^{ip\left (\frac {U}{2t}x-\sin k \right)}} {\cosh \frac {pU}{4t}} L_W(k)= \nonumber \\
&=& - \int _{-\pi}^{\pi} dk \, \cos k \, \delta (\sin k) \, L_{W_0}(k) + {\cal O}\left ( \frac{U}{t}\right)= -L_{W_0}(0)+L_{W_0}(\pi) + {\cal O}\left( \frac {U}{t}\right) \nonumber \\
&=& {\cal O}\left( \frac {U}{t} \right) \, ,
\end{eqnarray}
as follows from the form (\ref{W_0}) of $W_0(k)$. This allows us
to say that the solution of the NLIE for $Z$ in the limit $U/t
\rightarrow 0$ is ${\cal O}(U/t)$.

Stepping back to the $U/t$ expansion for $W$, the results on $Z$
allow to say that the term (\ref {zetazero}) is in fact ${\cal
O}(U^2/t^2)$. It follows that, at order $U/t$, the NLIE for $W$
reads
\begin{equation}
W_1(k)=\int _{-\pi}^{+\pi}dh \, \cos h \, \delta (\sin h - \sin k)
\,  {\mbox {Re}} \frac {e^{iW_0(h+i0)}}{1-e^{iW_0(h+i0)}} \,
W_1(h) \, ,
\end{equation}
whose solution is $W_1(k)=0$. Therefore, we can write that
\begin{equation}
W(k)=L(k-\phi)+2\pi N(k) + o\left(\frac{U}{t}\right) \, . \label
{W2ord}
\end{equation}

We are now ready to compute the leading term and its first
correction for the energy (\ref {Eexp}) of the highest energy
state in the limit $U/t\rightarrow 0$. For what concerns $E_L$,
Economou and Poulopoulos re-casted it \cite {EP} as an asymptotic
series in powers of $U/t$. We write only the first three terms of
such series:
\begin{equation}
E_L=-2t \left [ \frac {2L}{\pi} -\frac {UL}{8t}+ \frac {7 L \zeta
(3) U^2}{32 \pi ^3 t^2}+ {\cal O}\left(\frac{U^3}{t^3}\right)
\right ] \, . \label{BDS}
\end{equation}
We rewrite $E_Z$ after the change of variable $p\rightarrow pU/2t$
as
\begin{equation}
E_Z=2t \, \frac {U}{2t} \int _{-\infty}^{\infty} dx \left [ \int
_{-\infty}^{\infty} \frac {dp}{2\pi} \frac {\sin \frac
{pUx}{2t}}{\cosh \frac {pU}{4t}} J_1(p) \right ] L_Z(x) \, .
\end{equation}
Since $  {\mbox {Im}}\ln \left [1+e^{iZ(x+i0)}\right ]$ is ${\cal
O}(U/t)$ and the $p$-integral is also ${\cal O}(U/t)$, we conclude
that
\begin{equation}
E_Z=2t \, {\cal O}(U^3/t^3) \, . \label {Ezeta}
\end{equation}
We are left with the contributions coming from the third and the
fourth term of (\ref{Eexp}), $E_{W_1}$ and $E_{W_2}$, which we
rearrange, by reintroducing the function $G$, as follows
\begin{equation}
2t \int _{-\pi}^{\pi} \frac {dh}{\pi}\cos h\, L_W(h)\int
_{-\pi}^{+\pi} dk\ \frac {2t}{U} \sin k\ G\left [ \frac {2t}{U}
(\sin h -\sin k)\right] + 2t  \int _{-\pi}^{\pi} \frac
{dh}{\pi}\sin h\ L_W(h)  \, .
\end{equation}
Using (\ref {Gdelta}), we get
\begin{equation}
E_{W_1}+E_{W_2}=2t \left [ \int _{-\pi}^{\pi} \frac {dh}{\pi} \cos
h\ L_{W_0}(h)\frac {\sin h}{|\cos h|} + \int _{-\pi}^{\pi} \frac
{dh}{\pi}\sin h\  L_{W_0}(h) + o\left(\frac{U}{t}\right) \right ]
\, . \label {extra1}
\end{equation}
Putting together these two integrals, we are left with the
following addition to (\ref {BDS}):
\begin{equation}
E_{W_1}+E_{W_2}=4t \lim _{\epsilon \rightarrow 0} \left [ \int
_{-\frac {\pi}{2}}^{\frac {\pi}{2}} \frac {dh}{\pi}\sin h \ {\mbox
{Im}}\ln \left [1-e^{iW_0(h+i\epsilon)}\right ]+
o\left(\frac{U}{t}\right) \right ] \, . \label {extra2}
\end{equation}
After the insertion of (\ref {W_0}) in (\ref {extra2}), we are
left with the computation of
\begin{eqnarray}
&&4t \int _{-\frac {\pi}{2}}^{\frac {\pi}{2}} \frac {dh}{\pi}\sin
h \ {\mbox {Im}}\ln \left
[1-e^{iL(h-\phi+i\epsilon)}\right ]= \\
&&\frac {2t}{i} \int _{-\frac {\pi}{2}}^{\frac {\pi}{2}} \frac
{dh}{\pi}\sin h\ \ln \frac {1-e^{iL(h-\phi)-L\epsilon
}}{1-e^{-iL(h-\phi)-L\epsilon }}
  \, .
\end{eqnarray}
Integrating by parts gives
\begin{equation}
-\frac {2Lt}{\pi} \int _{-\frac {\pi}{2}}^{\frac {\pi}{2}} dh\
\cos h \left [\frac
{e^{iL(h-\phi)-L\epsilon}}{1-e^{iL(h-\phi)-L\epsilon}}+ \frac
    {e^{-iL(h-\phi)-L\epsilon}}{1-e^{-iL(h-\phi)-L\epsilon}}\right
]\, .
\end{equation}
In order to perform these integrations, we can see the ratios
involved as sums of geometric series
\begin{equation}
-\frac {2Lt}{\pi} \int _{-\frac {\pi}{2}}^{\frac {\pi}{2}} dh \cos
h \left [ \sum _{n=1}^{\infty}e^{iLn(h-\phi)-Ln\epsilon} +
 \sum _{n=1}^{\infty}e^{-iLn(h-\phi)-Ln\epsilon} \right ] \, .
\end{equation}
Now the integrations can be easily performed, giving
\begin{equation}
-\frac {2Lt}{\pi}\sum _{n=1}^{\infty}e^{-Ln\epsilon}\left [ \frac
{2\cos
      Ln\phi}{1-Ln}+ \frac {2\cos Ln\phi}{1+Ln} \right ]=
-\frac {8Lt}{\pi}\sum _{n=1}^{\infty}e^{-Ln\epsilon} \cos Ln\phi
\frac
      {1}{1-L^2n^2} \, .
\end{equation}
Going to the limit $\epsilon \rightarrow 0$ and rearranging this
series we get
\begin{equation}
E_{W_1}+E_{W_2}=2t \left [ \frac {4}{L\pi} \sum _{n=1}^{\infty}
\frac{\cos L n \phi}{n^2-\frac{1}{L^2}} +
o\left(\frac{U}{t}\right) \right]  \, .
\end{equation}
When $0\leq \phi <2\pi$ the sum of this series is (relation
1.445.6 of \cite{GR})
\begin{equation}
E_{W_1}+E_{W_2}=2t \left [ \frac {2L}{\pi}-2 \frac {\cos \left
(\frac {\pi}{L} - \phi \right)}{\sin \frac {\pi}{L}} +
o\left(\frac{U}{t}\right)  \right ]  \, . \label{sum}
\end{equation}
Summing (\ref {BDS}) with (\ref {sum}), we conclude that in the
limit $U/t\rightarrow 0$ the energy of the anti-ferromagnetic
state of the twisted Hubbard model behaves as
\begin{equation}
E=-2t \left [\frac {2\cos \left (\frac {\pi}{L} - \phi
\right)}{\sin \frac {\pi}{L}}  -\frac {UL}{8t} +
o\left(\frac{U}{t}\right) \right ] \, . \label{Efin}
\end{equation}
When $\phi =0$ we get the highest energy of the Hubbard model at
small coupling \cite {MV},
\begin{equation}
E=-2t \left [ 2\, {\mbox {cotan}} \frac {\pi}{L} -\frac {UL}{8t}+
o\left(\frac{U}{t}\right)  \right ] \, .
\end{equation}
On the other hand, according to \cite {RSS}, the Hamiltonian of
the twisted Hubbard model makes contact with the dilatation
operator of the $SU(2)$ sector of ${\cal N}=4$ SYM if $\phi =\pi
/2L$. In this case we get
\begin{equation}
E=-2t \left [ \frac {1}{\sin \frac {\pi}{2L}} -\frac {UL}{8t} +
o\left(\frac{U}{t}\right)  \right ] \, .
\end{equation}
We could go to higher orders, but this result already matches the
findings of \cite{BO} obtained by the usual first order
perturbation theory.

\subsubsection{On the strong coupling of the BDS Bethe Ansatz}
\label{strBDS} In \cite{FFGR} we proposed a NLIE for the BDS Bethe
Ansatz and we analysed in a careful detail the analytic form of
the finite size corrections. Because of the asymptotic nature of
such an Ansatz, we limited our discussion to the case when the
limit $L \to \infty$ is taken first, i.e. for finite $g$.

However, in the following we will compare the Hubbard model and
the BDS Ansatz for any range of $g$, and hence it is interesting
to return on the subject and discuss the structure of the finite
size corrections of the strong coupling limit of the NLIEs derived
in \cite{FFGR}, i.e. the residual $L$ dependence when the $g \to
\infty$ limit is taken first.

In the previous sections we have shown that the BDS Ansatz NLIE
can be formally obtained by those of the Hubbard model by simply
taking $W(k)=0$ \emph{tout court}. We can apply the same reasoning
here and derive the strong coupling behaviour of the energy for
the BDS Ansatz from the computation of the previous section.

Hence, by neglecting all those contributions coming from the
counting function $W(k)$ and collecting only those coming from
$Z(x)$, i.e. $E_L+E_Z$, we obtain
\begin{equation}
g\rightarrow \infty \,  \Rightarrow E_{\text {BDS}} =   \frac{ 2 L
\sqrt 2}{\pi \, g }   \, -\frac{ L}{4 \, g^2 } +\frac{7L\zeta
(3){\sqrt {2}}}{16 \pi ^3 g^3 } + {\cal O}\left (\frac {1}{g^4}
\right ) \, , \quad \forall L \, .  \label {EBDSg}
\end{equation}
Obviously, we have reintroduced the parametrization (\ref {tUg})
for the constants $t$ and $U$. It follows that the first three
terms of (\ref {EBDSg}), all coming from $E_L$, provide, up to
order ${\cal O}(g^{-3})$, the exact large $g$ limit of $E_{\text
{BDS}}$ for any $L$.

One can immediately realize that there is a stark difference
between the finite length corrections of this expression and those
obtained in \cite{FFGR}. We will return to this point in the next
section.


\section{Order of limits analysis}
\setcounter{equation}{0}

In order to achieve a satisfactory understanding of the behaviour
of the anomalous dimension for any value of the coupling constant,
it is important to analyse what happens to our equations when the
order of the limits $g,L \to \infty$ is changed. Such an aspect
can be addressed both in the twisted Hubbard model (with
$\phi=\pi/2L$) and in the BDS Ansatz, allowing us to compare them
explicitly.

It is important to stress that the NLIEs derived for those models
play a crucial role in order to have the sub-leading corrections
(in $g$ and $L$) under control. With this a piece of information
we will be able to infer some interesting properties about the
global behaviour of the anomalous dimension.

{\bf{The Hubbard model.}} \\
{\underline{$L,g \to \infty$}}: The analysis of Section
\ref{comparison} allows us to immediately write down the following
expression in the limit $L\to \infty$ and fixed $g$
\begin{equation}
L \to \infty \ \ \ \ \ \ E= \frac{\sqrt{2}\, L}{g} \int
_{-\infty}^{+\infty} \frac {dp}{p} \frac {J_0 (\sqrt2 g \, p ) J_1
(\sqrt2  g \, p ) }{e^{|p|}+1}+ \frac{\sqrt2}{L\pi g}\
\frac{I_1(\sqrt2 \pi g)}{I_0(\sqrt2 \pi g)}\ \frac{\pi^2}{6} \, +
\dots
\end{equation}
where we have an explicit expression of the $g$-dependent
coefficient of the $L$ and $1/L$ terms of the $L \to \infty$
expansion. A further expansion in $g$ gives
\begin{equation}
\label{H1} g \to \infty \ \ \ \ \ \ E^{(L,g)} = \frac{ 2 L \sqrt
2}{\pi \, g } +\frac{ \pi \, \sqrt 2}{6 \, L \, g }  \, -\frac{
L}{4 \, g^2 }
 \, + \dots
\end{equation}
{\underline{$g,L \to \infty$}}: Let us repeat the same
calculation, but reversing the order of the limits. Upon
exploiting one (known) result of Section \ref{strHubb}, we easily
conclude
\begin{equation}
\label{Hstg} g \to \infty \ \ \ \ \ \ E =  \frac{\sqrt 2}{g \;
\sin \frac{\pi}{2L}} \, -\frac{ L}{4 \, g^2 }   + \dots
\end{equation}
which gives the exact $L$-dependence of the $1/g$ term. By
expanding in $L$ we have
\begin{equation}
\label{H2} L \to \infty \ \ \ \ \ \ E^{(g,L)} = \frac{ 2 L \sqrt
2}{\pi \, g } +\frac12 \,\frac{ \pi \, \sqrt 2}{6 \, L \, g }  \,
-\frac{ L}{4 \, g^2 }  \, + \dots
\end{equation}
The conclusion is that, for the Hubbard model, the limits commute
\emph{only} at leading order in both $L$ and $g$. The disagreement
begins with the first sub-leading correction: it is interesting to
remark that when the order of the limits is exchanged, such a
correction conserves the same functional form and the only change
is in the numerical coefficient in front of it.

{\bf{The BDS Bethe Ansatz.}} \\
{\underline{$L,g \to \infty$}}: In ref.~\cite{FFGR} we computed
explicitly the large $L$ behaviour of the anomalous dimension
which turned out to be
\begin{equation}
L \to \infty \ \ \ \ \ \ E_{\text {BDS}} =\frac{\sqrt{2}\, L}{g}
\int _{-\infty}^{+\infty} \frac {dp}{p} \frac {J_0 (\sqrt2  g \, p
) J_1 (\sqrt2 g \, p ) }{e^{|p|}+1}+ \frac{\sqrt2}{L\pi g}\
\frac{I_1(\sqrt2 \pi g)}{I_0(\sqrt2 \pi g)}\ \frac{\pi^2}{6} \, +
\dots
\end{equation}
As explained in Section \ref{comparison}, when $L\rightarrow
\infty$ the highest energies of BDS and Hubbard models coincide,
up to exponentially small terms. Hence, expanding in $g$, we have
again
\begin{equation}
\label{BDS1} g \to \infty \ \ \ \ \ \ E_{\text {BDS}}^{(L,g)}=
\frac{ 2 L \sqrt 2}{\pi \, g } +\frac{ \pi \, \sqrt 2}{6 \, L \, g
}  \,
 \, -\frac{ L}{4 \, g^2 } + \dots
\end{equation}
{\underline{$g,L \to \infty$}}: This case was discussed in Section
\ref{strBDS}. The $g\to \infty$ limit gives
\begin{equation}
g \to \infty \ \ \ \ \ \ E_{\text {BDS}} =   \frac{ 2 L \sqrt
2}{\pi \, g }   \, -\frac{ L}{4 \, g^2 } +\frac{7L\zeta (3){\sqrt
{2}}}{16 \pi ^3 g^3 } +
 \dots
\end{equation}
Therefore, we conclude that
\begin{equation}
\label{BDS2} L \to \infty \ \ \ \ \ \ E_{\text {BDS}}^{(g,L)}=
\frac{ 2 L \sqrt 2}{\pi \, g }  \, -\frac{ L}{4 \, g^2 } +
\frac{7L\zeta (3){\sqrt {2}}}{16 \pi ^3 g^3 } +
 \dots
\end{equation}
As expected, in the BDS Bethe Ansatz the order of the limits
commutes only at leading order in $g$ and $L$. It is important to
point out that the sub-leading corrections differs also in their
functional form, because the term which behaves as $1/ (g \, L)$
is absent. \vskip0.4cm {\bf{Remarks.}}
\begin{enumerate}
\item{The discussion of this section has many points of contact
with that of Section 3 of \cite{BO2}. The main difference is
related to the treatment of the sub-leading corrections in $L$ for
$ E_{\text {BDS}}^{(g,L)}$. If one takes the $g \to \infty$ limit
starting from the NLIEs of \cite{FFGR}, one immediately realizes
that the sub-leading correction used in ref. \cite{BO2} (given by
equation (\ref{BDS1})) is not the correct one, because in this
limit the structure of the counting function changes dramatically
giving the structure observed in (\ref{BDS2}). In particular our
analysis of Section \ref{strBDS} shows that sub-leading
corrections in $1/L$ will appear only beyond the order $1/g^3$.}
\item{As expected from the results of Section \ref{comparison}, if
we take first the limit $L \to \infty$, the Hubbard and BDS Ansatz
behave the same way. This is because, in such a limit the charge
degrees of freedom (described by the counting function $W$ in the
language of the NLIEs) are exponentially depressed.} \item{In the
BDS Ansatz it was somehow expected that the limits do not commute,
in particular because of the so-called "wrapping problem". What is
more surprising is that the limits do not commute also in the
Hubbard model which is believed not to be plagued by such a
pathology.} \item{As previously pointed out in \cite{BO2}, the
leading strong coupling behaviour at infinite length is the same,
no matter what the model and order of limits are considered.}
\end{enumerate}
In the following sections about the numerical analysis at fixed
$L$ we will use the following expressions for the strong coupling
expansion
\begin{eqnarray}
\label{strcoup}
E & = &  \frac{\sqrt 2}{g \; \sin \frac{\pi}{2L}} \, -\frac{ L}{4 \, g^2 }   + \dots  \nonumber \\
E_{\text {BDS}} & = & \frac{ 2 L \sqrt 2}{\pi \, g }  \, -\frac{
L}{4 \, g^2 } + \dots.
\end{eqnarray}


\section{Numerical analysis}
\setcounter{equation}{0}

The equations obtained in the previous sections are suitable for
numerical evaluations, and, as in \cite{FFGR}, it is not difficult
to solve them by iteration. Our analysis is mainly devoted to
investigate the difference between the highest energies (or
anomalous dimensions) in the Hubbard model and in the BDS Bethe
Ansatz. For this reason, we will uniquely use the coupling $g$ and
we will use (\ref{tUg}) to express $t$ and $U$. In particular, we
observe the value of the ratio
\begin{equation}
\frac{t}{U}=\frac{g}{\sqrt{2}} \, .
\end{equation}
For BDS, calculations were also performed in \cite{FFGR}. We
remember that this model is conjectured to work only when $g\ll 1$
and up to the order $g^{2L-2}$, after which the wrapping problem
is present. In spite of this we will compare the energies of
Hubbard and BDS predictions out of the strictly perturbative
regime. We will begin with the analysis of the Konishi operator
($L=4$), then we will study the behaviour of the highest energy
state in the case of chains with an intermediate number of sites.
We conclude with a numerical study of the difference between the
BDS Ansatz and the Hubbard model for large $L$ and finite $g$ in
order to provide a numerical support to the analytic results of
Section \ref{comparison}.

\subsection{A test for the NLIEs: the Konishi operator}

\begin{figure}[h]
\setlength{\unitlength}{1mm}
\begin{picture}(100,104)(-3,0)
\put(0,0){\includegraphics[width=0.95\linewidth]{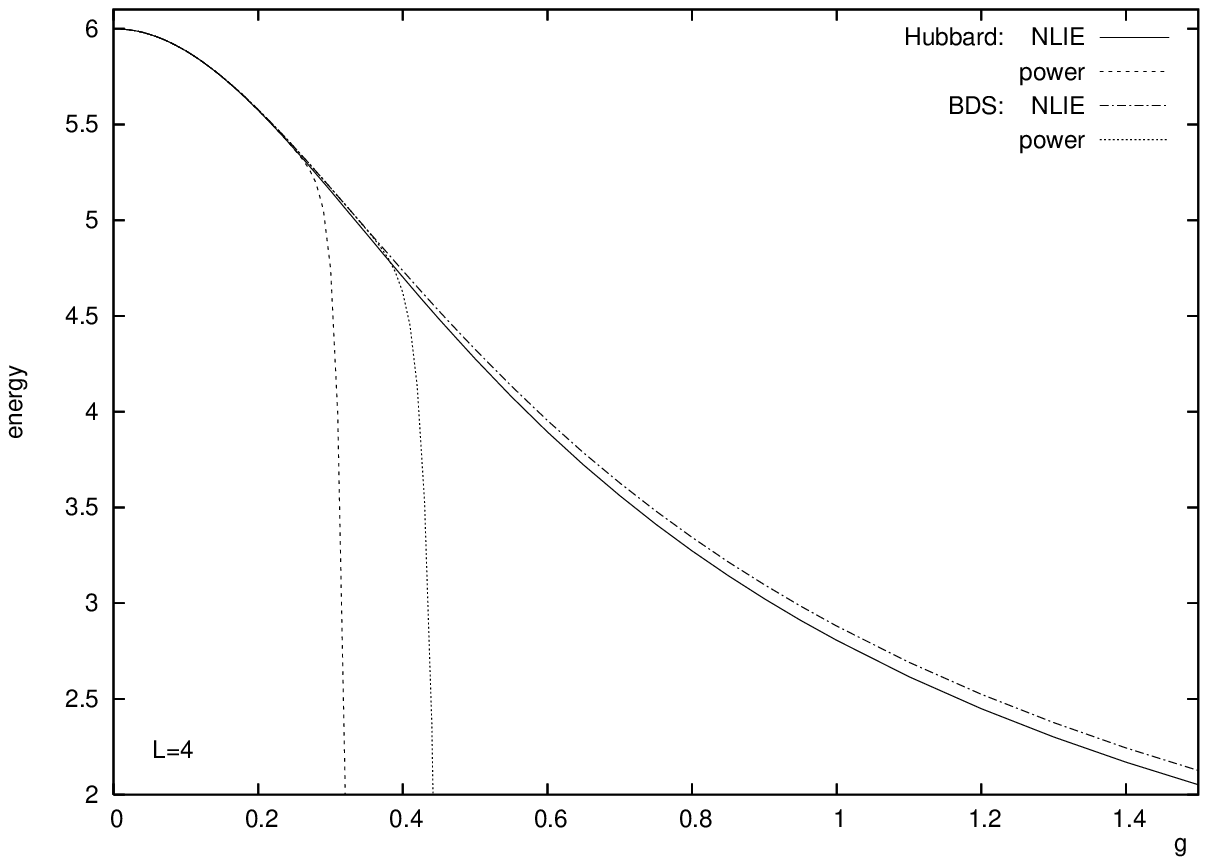}}
\put(139,40){\color{white}\rule{4.5mm}{40mm}}
\put(79,35){\includegraphics{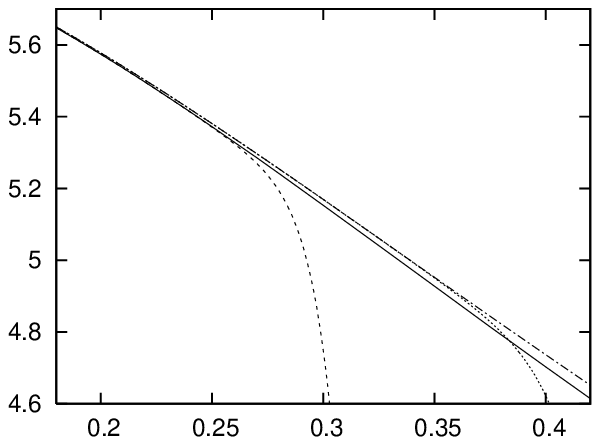}}
\put(30,66){\drawline(0,0)(21,0)(21,25)(0,25)(0,0)}
\put(50.8,91){\dottedline{3}(0,0)(38.4,-11)}
\put(50.8,66){\dottedline{3}(0,0)(38.4,-26)}
\end{picture}
\caption{\label{konishi1} Comparison of the highest Hubbard and
BDS energies (anomalous dimensions) for a system with $L=4$ sites,
corresponding to the anomalous dimension of the Konishi operator.
The curves indicated by ``NLIE'' are obtained by the non-linear
integral equation -- (\ref{Eexp}) for Hubbard and eq. 3.24 of
\cite{FFGR} for BDS -- and those indicated by ``power'' are
obtained with the power expansions (\ref{powexpH} and
\ref{powexpBDS}) up to the thirtieth order. The value of $g$ where
they get far away gives an idea of the convergence radius. The
small image is a zoom of the surrounded area of the largest one.}
\end{figure}
\begin{figure}[h]\hspace*{21mm}
\includegraphics[width=0.7\linewidth]{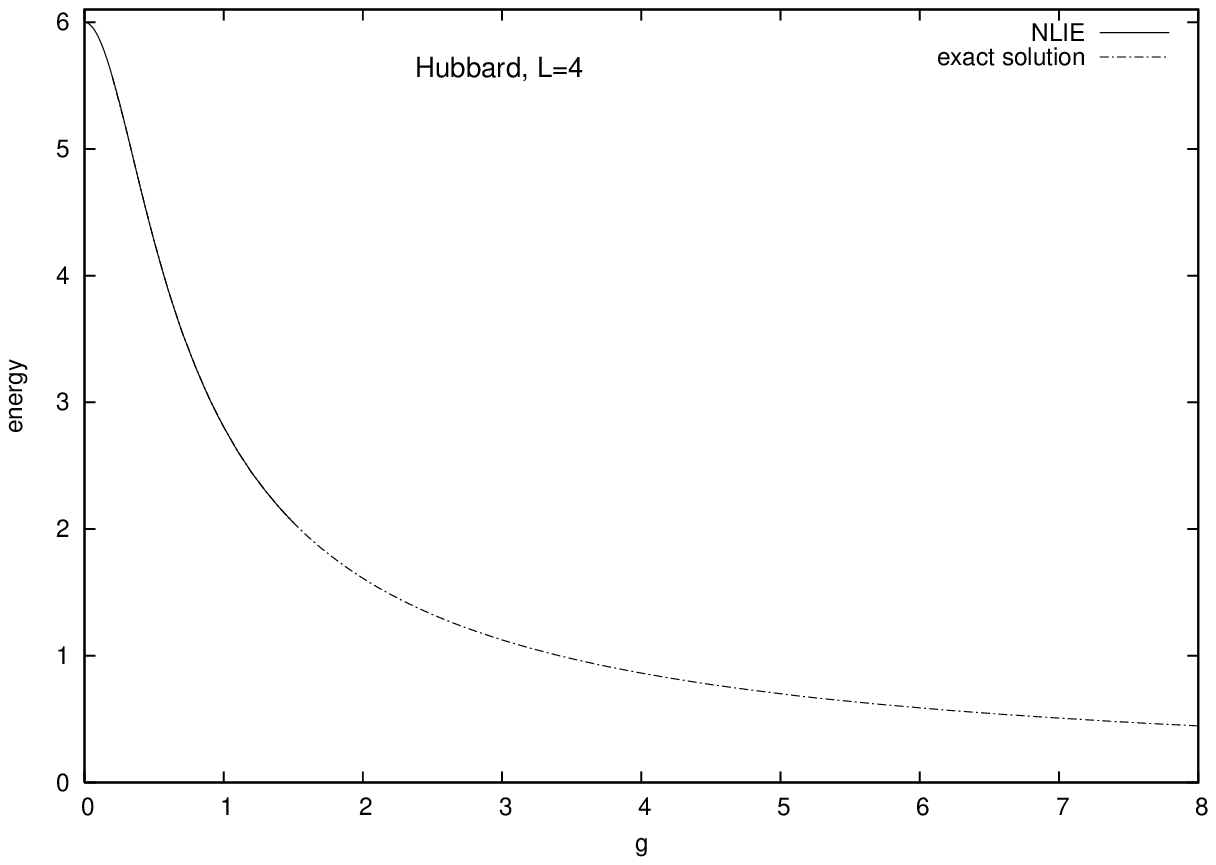}
\caption{\label{nliemin} Comparison of numerical NLIE data (the
same as in Fig.~\ref{konishi1}) and the exact albeit implicit
solution of \cite{MIN}, from weak to strong coupling. The two
curves perfectly overlap in the common interval $g\in[0,1.5]$.}
\end{figure}

The highest energy state with $L=4$, corresponding to the Konishi
operator, has been extensively studied from the perturbative point
of view, both in the context of the BDS Ansatz and within the
Hubbard model. Furthermore, for the latter, an exact implicit
expression for the anomalous dimension has been found by Minahan
\cite{MIN}.

In the present section we will use such known results as a test
for the validity of the NLIEs derived in this paper.

Let us begin with the comparison between our NLIE and the
perturbative expansions calculated in \cite{RSS} (for our
convenience we pushed the computation up to the thirtieth order by
using the \textsf{Mathematica} routines in Appendix~B of
\cite{RSS}, see below).

The result of such a comparison is summarized in
Fig.~\ref{konishi1}. Firstly, it is impressing the behaviour of
the perturbative expansion when compared to the exact result
coming from the NLIEs: the convergence seems to be quite slow and
the perturbative window turns out to be very small. We can also
observe in Fig.~\ref{konishi1} that the first line which separates
is the one corresponding to (\ref{powexpH}) and the second is
(\ref{powexpBDS}). In the small image we have magnified the region
where the perturbative expansions and the exact curves separate.

The emergence of such a behaviour can be explained by the
appearance of rapidly growing coefficients in both  the Hubbard
model and BDS Ansatz cases \footnote{Note that the Hubbard
coefficients are (multiple of $6$) integers and so are the BDS
ones if $g$ is properly re-scaled.}
\begin{eqnarray}
E &=& 6-12\ g^2 +42\ g^4 - 318\ g^6+ 4524\ g^8-63786\
g^{10}+783924\ g^{12}
-\nonumber \\
&& -8728086\ g^{14} +93893622\ g^{16}-1038217494\ g^{18} +
12181236666\ g^{20}
-\nonumber\\
&&-150141359712\ g^{22}  + 1888713236976\ g^{24} - 23751656065164\ g^{26} \nonumber\\
&& +297019282258320\ g^{28}
 - 3710023076959086\ g^{30}+\dots \,,
 \label{powexpH} \\
E_{\text{BDS}} &=& 6-12\ g^{2} +42\ g^{4} -\frac{705}{4}\
g^{6}+\frac{6627}{8} g^{8} -\frac{67287}{16}\ g^{10}
+\frac{359655}{16}\ g^{12}-\nonumber \\ && -\frac{7964283}{64}\
g^{14} +\frac{22613385}{32}\ g^{16} - \frac{261928101}{64}\ g^{18}
+\frac{6164759913}{256}\ g^{20}- \nonumber \\&& -
\frac{147007778043}{1024}\  g^{22} + \frac{1772167996011}{2048} \
g^{24} -\frac{10781715497325}{2048}\ g^{26}
\nonumber \\
&& +\frac{66122074282395}{2048}\ g^{28}
-\frac{3266715687275811}{16384}\ g^{30}+\dots \, .
\label{powexpBDS}
\end{eqnarray}

Another interesting check is given by the comparison of our
numerical results with the exact (implicit) form for the anomalous
dimension of the Konishi operator given by Minahan in \cite{MIN}.
As shown in Fig.~\ref{nliemin}, we found a complete agreement
within the numerical precision of our computation. This allowed us
to estimate our relative error to be less than $2\cdot 10^{-6}$.

It is interesting to notice that, even if $L$ is not at all large,
the difference between the exact curves for Hubbard and BDS
remains small as $g$ is increased. This fact seems to suggest that
the BDS Ansatz can be considered as a good approximation of the
Hubbard model even when $g$ is large (i.e. non-perturbative) and
$L$ is fixed to a small value (and not only in the limit of
infinite chain).

From this point of view it would be nice to study what happens in
the strong coupling regime. Unfortunately such a regime is
difficult to reach with the numerical integration of our NLIEs,
because of a reduced numerical precision for large $g$.

However, it is possible to use the strong coupling expansions eq.
(\ref{strcoup}) for the present $L=4$ case. We obtain
\begin{eqnarray}
E &=& \frac{3.69552}{g}-\frac{1}{g^2}+\dots \\
E_{\text {BDS}} &=& \frac{3.60127}{g}-\frac{1}{g^2}+\dots.
\end{eqnarray}
This result is interesting for two reasons. Firstly, it shows that
already for $L=4$ the strong coupling prediction for the BDS
Ansatz is a good approximation of the corresponding result
obtained in the Hubbard model. Moreover, one can see that the
strong coupling expansions smoothly joins the numerical data in
Fig.~\ref{konishi1}: we will further comment on this issue in the
next section where we will analyse operators of intermediate
length.

\subsection{Intermediate length operators}

Since our equations are suitable to the study of the energy at any
$L$, we can use them to analyze operators of intermediate length.
This is the case where the NLIEs can be exploited at their best,
because in such an intermediate regime it is quite difficult to
directly use the Bethe Ansatz equations due to the large number of
terms. In particular we choose to analyze highest energy states
with length $L=12$ and $L=40$.

\begin{figure}[h]\hspace*{12mm}
\includegraphics[width=0.8\linewidth]{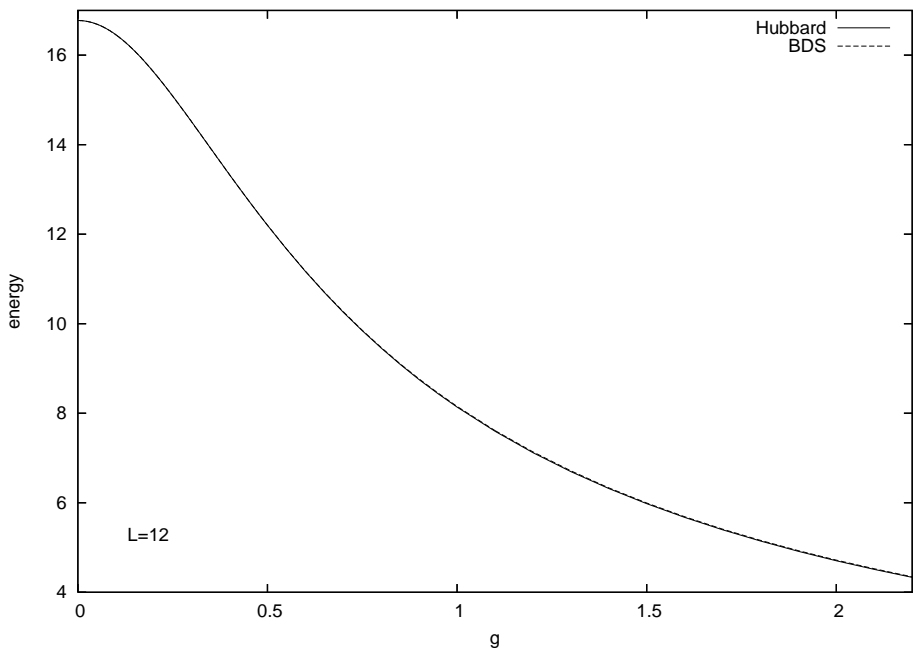}
\caption{\label{antif12} Comparison of Hubbard and BDS energies
(anomalous dimensions) for \mbox{$L=12$}. The two curves are
almost indistinguishable in this range of $g$ and start to
separate at the right border of the plot.}
\end{figure}
\begin{figure}[h]\hspace*{3mm}
\includegraphics[width=0.95\linewidth]{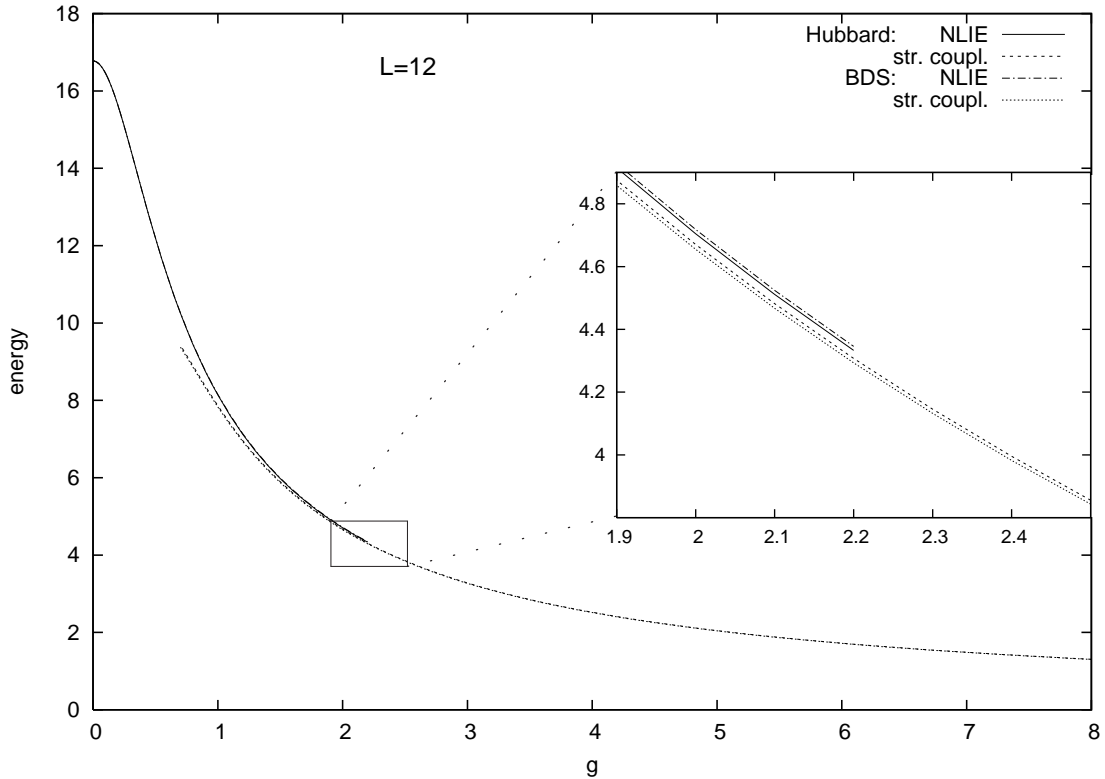}
\caption{\label{ws} The behaviour of the energies $E(g)$ and
$E_{\text {BDS}} $ from small to strong coupling is plotted here
for a lattice of 12 sites. The left branches of the curves are the
same as in Fig.~\ref{antif12} while the right branches are given
by the strong coupling expansions (\ref{strong}). In the small
picture there is a zoom of the region where the branches overlap.}
\end{figure}
\begin{figure}[h]\hspace*{12mm}
\includegraphics[width=0.8\linewidth]{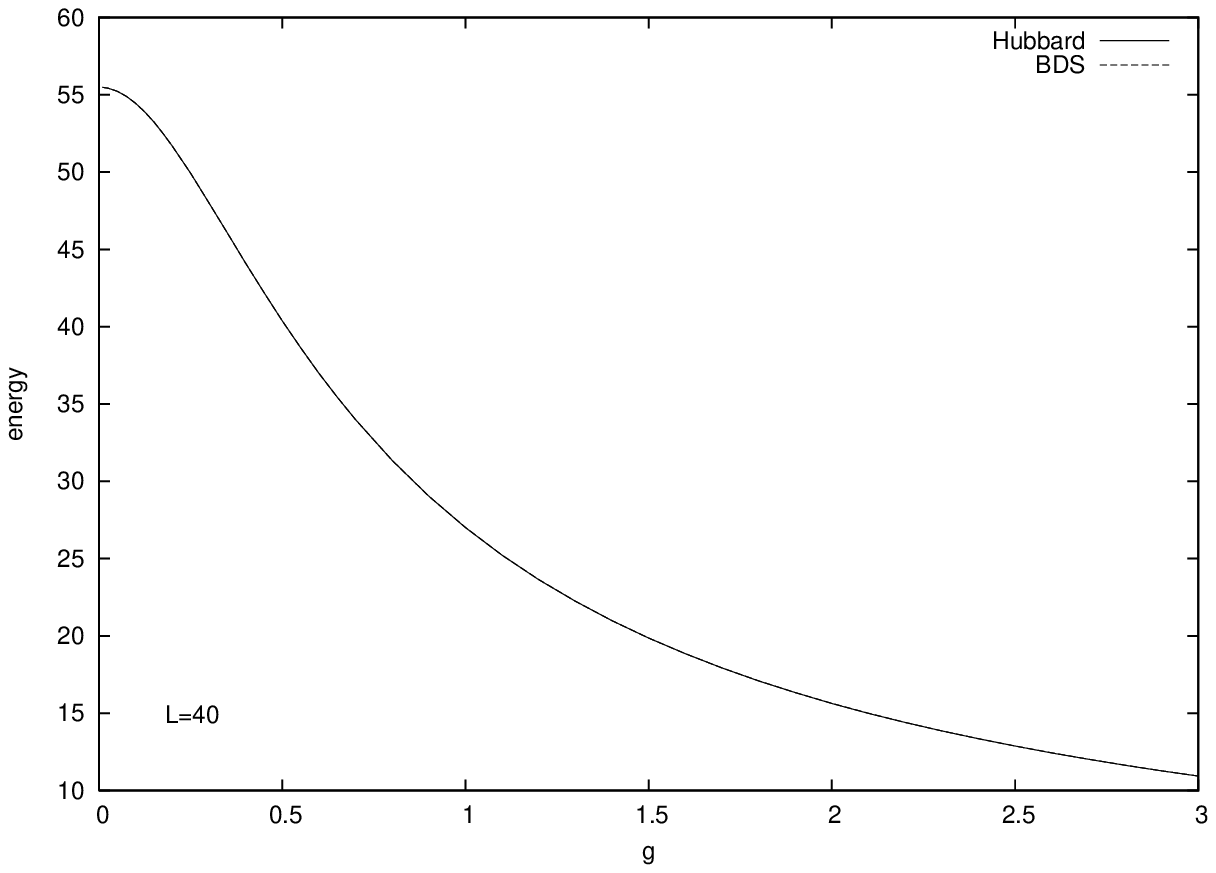}
\caption{\label{antif40} Comparison of Hubbard and BDS energies
(anomalous dimensions) for $L=40$. The two curves are visibly hard
to distinguish in this range of $g$. The largest reached absolute
difference is 0.0038 namely a relative difference of $0.019\%$.}
\end{figure}
Let us begin with $L=12$. Figure \ref{antif12} shows that already
at such a small value of $L$, and beyond the perturbative region
in the coupling $g$, the curves computed by means of both BDS
Ansatz and Hubbard model overlap almost completely. This is an
indication that even if we are dealing with a chain  which is far
from the thermodynamic limit, the predictions of the BDS Ansatz
can be considered as a good effective approximation of the Hubbard
model behaviour.

Hence, even if the BDS Ansatz is plagued by the wrapping problem,
at a quantitative level it is able to reproduce all the
significant features of the Hubbard model, beginning from $L
\simeq 12$.

Again, we use eqs. (\ref{strcoup}) to describe the strong coupling
behaviour
\begin{eqnarray}
E & = & \frac{10.8347}{g} - \frac{3}{g^2}
+\dots \nonumber \\
E_{\text {BDS}} & = & \frac{10.8038}{g} - \frac{3}{g^2} +\dots .
\label{strong}
\end{eqnarray}
The behaviour from weak to strong coupling for both the Hubbard
model and the BDS anstaz is plotted in Fig.~\ref{ws}: the left
branches are the same as in Fig.~\ref{antif12} (numerical solution
of the NLIEs), while the right branches are given by the equations
(\ref{strong}) (strong coupling expansions). As we stated at the
end of previous subsection, left and right branches smoothly join.

Let us remark that it was quite unexpected to find the observed
good agreement between the Hubbard model and the BDS Ansatz
predictions for such a small value of $L$. It is also important to
stress the crucial role played by the NLIEs in order to obtain the
exact behaviour of the energy outside the perturbative domain. As
shown in the study of the anomalous dimension of the Konishi
operator, the perturbation theory alone is not enough to reach an
overlap with the strong coupling expansion.

Finally, we compared the Hubbard and BDS anomalous dimensions for
$L=40$ in the range $g \in [0,3]$: as expected, the agreement
between them is further enhanced and the two curves can be hardly
distinguished, see Fig.~\ref{antif40}.

The interesting feature of this case is that we were able to
explicitly follow the evolution of the relative difference between
Hubbard and BDS from weak to strong coupling. We observed that the
two curves begin to separate at small $g$, then they achieve a
maximum in the relative difference, and after that they start to
approach again. We think that such a pattern is valid at any $L$,
but the reduced numerical precision does not allow us to observe
it for smaller values of $L$.

{\bf Remark}. The fact that $E_{\text {BDS}}>E$ at weak coupling,
but $E_{\text {BDS}}<E$ at strong coupling suggests that the
curves will cross at some intermediate value of $g$. This is
consistent with our numerical observation of the approaching of
them as $g$ increases. Unfortunately, the reduced precision of our
data at large $g$ prevented us to observe such a crossing
explicitly.

\subsection{Large operators at fixed $g$\label{loafg}}
\begin{table}[h]
\caption{\label{tabella} Table of data concerning the exponential
damping $|E_{\text{BDS}}-E|\propto e^{-a(g)L}$ as in
(\ref{expdamp2}) and in Section~\ref{comparison}. Two significant
digits are available for the data obtained by numerical
integration of the NLIEs namely for $a(g)$. The inequality
(\ref{ineq}) is satisfied with a slowly changing or possibly
constant ratio.}
\begin{equation*}
\begin{array}{r|ccc}
g & 1.2 & 1.6 & 2 \\
\hline
a(g) & 0.0076 &0.0011 & 0.00017\\
\omega(g) & 0.0174422 & 0.00342347 & 0.000649349 \\
\epsilon_M  & 0.290524 & 0.219211 & 0.175869\\
\epsilon_M \omega(g)& 0.00506738 &0.000750462 & 0.000114200 \\
\frac{a(g)}{\epsilon_M\, \omega(g)} &1.5 &1.5 & 1.5
\end{array}
\end{equation*}
\end{table}
\begin{figure}[h]
\includegraphics[width=0.49\linewidth]{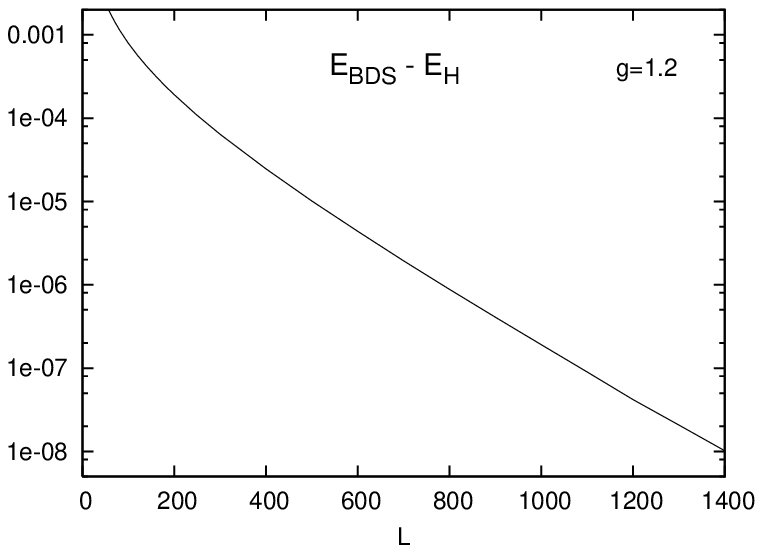}
\includegraphics[width=0.49\linewidth]{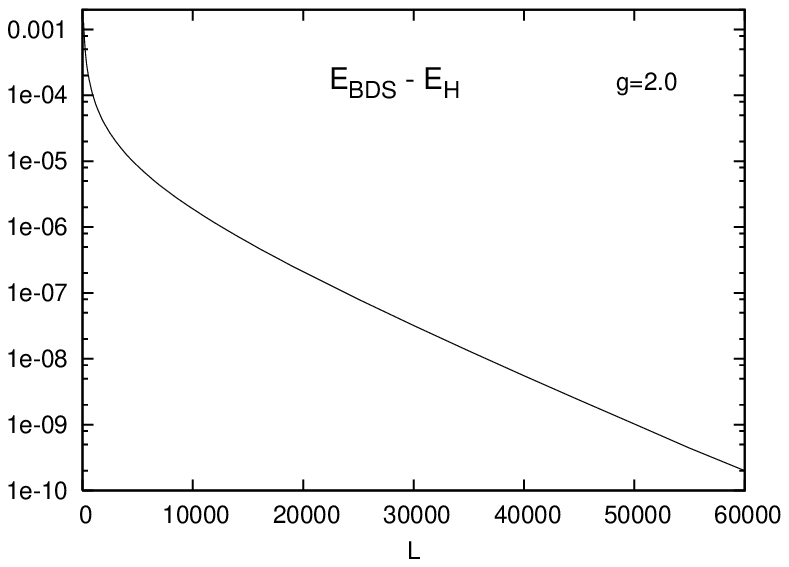}
\caption{\label{relative} Difference between the highest Hubbard
and BDS energies (anomalous dimensions) at different sizes of the
system and at two fixed values of the coupling, \mbox{$g=1.2$} and
$g=2$. A logarithmic scale is used on the vertical axis. The
linear behaviour is clear indication of the exponential damping
discussed in Section~\ref{comparison} and summarized in
(\ref{expdamp2}). Related numerical data are provided in
Table~\ref{tabella}.}
\end{figure}

Another interesting analysis concerns the difference between the
highest energies of Hubbard and BDS as a function of $L$ and at
fixed coupling. Our choices were the values $g=1.2,~1.6$ and
$g=2$, which lie in an intermediate region for which our
asymptotic result (\ref {lglimit}) does not apply. The solution of
our NLIEs gives the result depicted in Table~\ref{tabella} and in
Fig.~\ref{relative} on a log diagram: for large $L$ ($L>300$) the
behaviour is linear, meaning that the difference between the
energies decays exponentially as the length $L$ is increased. This
confirms our analytical findings of Section~\ref{comparison}.
Consistently with that Section, we introduce the numerical rate of
decay $a(g)$ and write
\begin{equation}\label{expdamp2}
|E_{\text{BDS}}-E| \propto e^{-a(g) L}\,, \qquad a(g)>0 \,.
\end{equation}
According to our results of Section 5, the inequality
\begin{equation}\label{ineq}
\epsilon\, \omega(g) \leq \epsilon_M \, \omega(g) \leq a(g) \,
\end{equation}
must hold and in this respect, Table~\ref{tabella} suggests that
(\ref{ineq}) is actually correct for the values of $g$ chosen.

The values of $a(g)$ shown in Table~\ref{tabella} interpolate
between the behaviour at small $g$, $a(g)=-2\ln g$, and at large
$g$, $a(g)=\frac {1}{\sqrt {2}g}$. Consistently, the values of
$a(g)$ in Table~\ref {tabella} decrease as $g$ increases. However,
understanding how $a(g)$ passes from the small coupling to the
large coupling behaviour on the basis of numerical data seems
difficult. Indeed, the comparison of the two plots in
Fig.~\ref{relative} shows that the exponential behaviour in $L$ of
$E-E_{\text {BDS}}$ is strongly dependent on the actual value of
the coupling. In addition, it is clear that linearity (i.e.
exponential damping) is reached at values of $L$ which rapidly
increase with $g$.

\begin{figure}[h]
\includegraphics[width=0.49\linewidth]{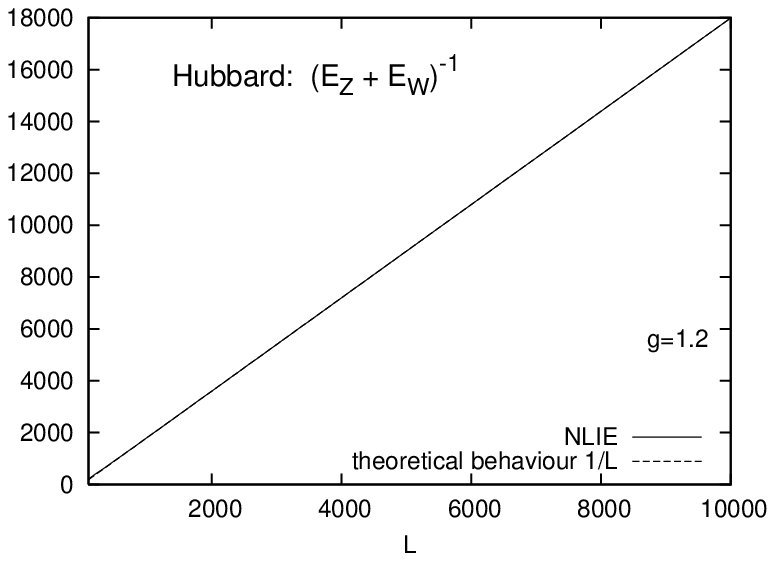}
\includegraphics[width=0.49\linewidth]{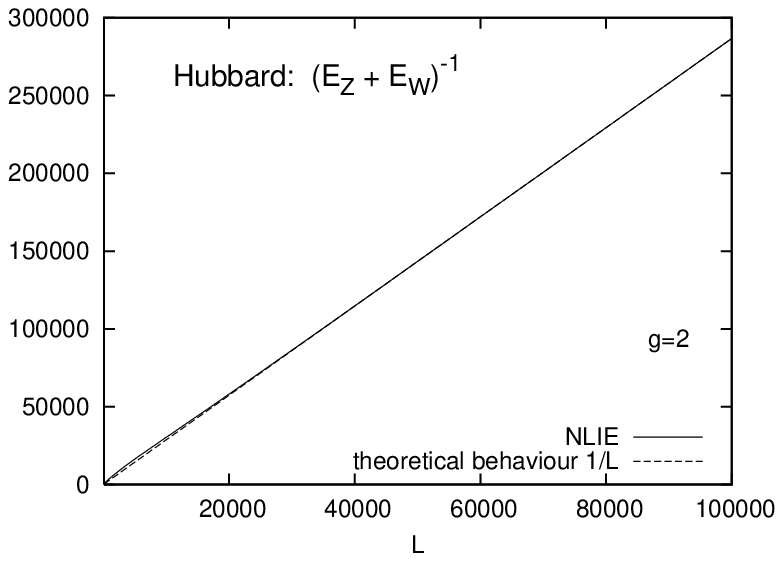}
\caption{\label{unosul} The finite size correction $1/L$ is
compared with the theoretical BDS prediction (\ref{fsc}),
represented with the dashed straight line. As observed in
Fig.~\ref{relative}, for higher values of $g$ linearity is reached
at higher values of L.}
\end{figure}
Finally, we remember, as pointed out before equation (\ref{fsc}),
that the exponential damping of $E-E_{\text {BDS}}$ forces the
equality of logarithmic-like and power-like finite size
corrections in the Hubbard and BDS models. The most relevant case
is the $1/L$ behaviour, that is explicitly plotted for the Hubbard
model in Fig.~\ref{unosul} and compared with the analytical
prediction (\ref{fsc}) coming from the BDS Ansatz.

\section{Summary and perspective}

In this paper we have derived the NLIEs describing the highest
energy state of the half-filled (attractive) Hubbard model with
and without a suitable flux which is responsible for a precise
contact with the highest possible anomalous dimension (at fixed
bare dimension $L$). In particular, according to the important
correspondence pioneered in \cite{RSS}, we computed the
energy/anomalous dimension for this state/operator, and all the
other conserved quantity eigenvalues may find an exact expression.
The dimension is of course a function of the 't Hooft coupling
$\lambda$ (and of the operator bare dimension $L$), thanks to a
duality connexion between this latter and the Hubbard coupling.
While exact analytical formul\ae \, for energy/dimension have been
extracted only in the strong and weak coupling perturbative
regime, numerical solutions of the NLIEs and corresponding energy
values can be obtained for arbitrary values\footnote{Despite this,
we experienced an increasing numerical error while increasing
$g$.} of $g$ and $L$. In this respect, we have been able to
provide many plots showing clearly the dependence of the highest
energy (anomalous dimension) on the coupling $g$. In particular,
we have concentrated ourselves on the comparison with the highest
energy of the (simpler) BDS model. We have shown, first
analytically then numerically, that at large sizes ($L\rightarrow
\infty$), they coincide up to corrections exponentially small
(i.e. $o(L^{-\infty}$)), i.e. of non-analytic form. Of course, the
finite size corrections are also important in the string theory
since they come out as quantum loop corrections. As regards future
perspectives, we would like to stress that the NLIE approach is a
very good and effective method to provide the observables
dependence on the model parameters and size. One good quality is
that the NLIE can be written whenever the Bethe Ansatz equations
are available, and even under more general circumstances, thanks
to its equivalence to some functional equations. In particular the
application of the NLIE to the so-called string Bethe Ansatz
\cite{BES} and larger field sectors is particularly desirable, in
view of tests of the AdS/CFT correspondence. For the time being,
indeed, in spite of the progress following the discovery of
integrability in both sides of the duality, such tests could be
performed only for a limited number of cases, because of the
technical difficulty in handling the Bethe equations for general
length and coupling. In the previous work \cite {FFGR} and in the
present one we have showed that by means of the NLIE framework we
are able to overcome this difficulty and provide exact analytic
scaling with the length $L$ and the coupling $\lambda$ (and also
numerical evaluation of the conformal dimensions).

\section*{Acknowledgments}

We have the pleasure to acknowledge useful discussions with D.
Bombardelli, A. Cappelli, A. Doikou, E. Ercolessi, G. Ferretti, A.
Montorsi, F. Ravanini, M. Staudacher and K. Zarembo; V. Rittenberg
enthusiastic support was, besides, invaluable. We are all indebted
to EUCLID, the EC FP5 Network with contract number
HPRN-CT-2002-00325, which, in particular, has supported the work
of PG. GF thanks INFN for a post-doctoral fellowship. MR thanks
the INFN and the Department of Physics in Bologna for warm
hospitality and support. DF thanks the INFN (especially grant {\it
Iniziativa specifica TO12}) for travel and invitation financial
support.




\begin{thebibliography}{99}


\bibitem{KW}
A. Kapustin, E. Witten, {\sl Electric-Magnetic Duality and the
Geometric Langlands Program}, hep-th/0604151;


\bibitem{MWGKP}
J.M. Maldacena, {\sl The large N limit of superconformal field
theories and supergravity}, Adv. Theor. Math. Phys.
{\bf 2} (1998) 231 and hep-th/9711200; \\
E. Witten, {\sl Anti-de Sitter space and holography}, Adv. Theor.
Math. Phys. {\bf 2} (1998) 253 and
hep-th/9802150; \\
S.S. Gubser, I.R. Klebanov, A.M. Polyakov, {\sl Gauge theory
correlators from non-critical string theory}, Phys.Lett. {\bf
B428} (1998) 105 and hep-th/9802109;

\bibitem{MZ}
J.A. Minahan, K. Zarembo, {\sl The Bethe Ansatz for ${\cal N}=4$
Super Yang-Mills}, JHEP{\bf 03} (2003) 013 and hep-th/0212208;

\bibitem{Bethe}
H. Bethe, {\sl On the theory of metals, 1. Eigenvalues and
eigenfunctions for the linear atomic chain}, Z. Phys. {\bf 71}
(1931) 205;

\bibitem{HLPS}
R. Hernandez, E. Lopez, A. Perianez, G. Sierra, {\sl Finite size
effects in ferromagnetic spin chains and quantum corrections to
classical strings}, JHEP{\bf 06} (2005) 011 and hep-th/0502188;

\bibitem{BTZ}
N. Beisert, A.A. Tseytlin, K. Zarembo, {\sl Matching quantum
strings to quantum spins: one-loop vs. finite size corrections},
Nucl. Phys. {\bf B715} (2005) 190 and hep-th/0502173;

\bibitem{GK}
N. Gromov, V. Kazakov, {\sl Double scaling and Finite Size
Corrections in $sl(2)$ Spin Chain}, Nucl. Phys. {\bf B736} (2006)
224 and hep-th/0510194;

\bibitem{BMN}
D. Berenstein, J.M. Maldacena, H. Nastase, {\sl Strings in flat
space and pp waves from ${\cal N}=4$ super Yang-Mills}, JHEP {\bf
04} (2002) 013 and hep-th/0202021;

\bibitem{BKS}
N. Beisert, C. Kristjansen, M. Staudacher, {\sl The dilatation
operator of ${\cal N}=4$ super Yang-Mills theory}, Nucl. Phys.
{\bf B664} (2003) 131 and hep-th/0303060;

\bibitem{SS}
D. Serban, M. Staudacher, {\sl Planar ${\cal N}=4$ gauge theory
and the Inozemtsev long range spin chain}, JHEP{\bf 06} (2004) 001
and hep-th/0401057;

\bibitem{BDS}
N. Beisert, V. Dippel, M. Staudacher, {\sl A novel long range spin
chain and planar ${\cal N}=4$ super Yang-Mills}, JHEP{\bf 07}
(2004) 075 and hep-th/0405001;

\bibitem{RSS}
A. Rej, D. Serban, M. Staudacher, {\sl Planar ${\cal N}=4$ gauge
theory and the Hubbard model}, JHEP{\bf 03} (2006) 018 and
hep-th/0512077;

\bibitem{MIN}
J.A. Minahan, {\sl Strong coupling from the Hubbard model},
J.Phys.A{\bf 39} (2006) 13083 and hep-th/0603175;

\bibitem{BES}
N. Beisert, B. Eden, M. Staudacher, {\sl Trascendentality and
Crossing}, hep-th/0610251;

\bibitem{LW}
E.H. Lieb, F.Y. Wu, {\sl Absence of Mott-transition in an exact
  solution of the short-range, one-band model in one dimension},
Phys. Rev. Lett. {\bf 20} (1968) 1445; \\
E.H. Lieb, F.Y. Wu, {\sl The one-dimensional Hubbard model: A
  reminiscence}, cond-mat/0207529;



\bibitem{FFGR}
G.Feverati, D.Fioravanti, P.Grinza, M.Rossi, {\sl On the finite
size corrections of anti-ferromagnetic anomalous dimensions in
${\cal N}=4$ SYM}, JHEP{\bf 05} (2006) 068 and hep-th/0602189;


\bibitem{KP}
P.A. Pearce, A. Kl\"umper, {\sl Finite-size corrections and
scaling dimensions of solvable lattice models: an analytic
method}, Phys. Rev. Lett.  {\bf 66}, volume 8 (1991) 974;\\
A. Kl\"umper, M.T. Batchelor and P.A. Pearce, {\sl Central charges
of the 6- and 19-vertex models with twisted boundary conditions},
J. Phys. {\bf A24} (1991) 3111;

\bibitem{DDV}
C. Destri, H.J. de Vega, {\sl New thermodynamic Bethe Ansatz
equations without strings}, Phys. Rev. Lett. {\bf 69} (1992) 2313;\\
C. Destri, H.J. de Vega, {\sl Unified approach to Thermodynamic
Bethe Ansatz and finite size corrections for lattice models and
field theories}, Nucl. Phys. {\bf B438} (1995) 413 and
hep-th/9407117;

\bibitem{FMQR}
D. Fioravanti, A. Mariottini, E. Quattrini, F. Ravanini, {\sl
Excited state Destri-de Vega equation for sine-Gordon and
restricted sine-Gordon models}, Phys. Lett. {\bf B390} (1997) 243
and hep-th/9608091;\\
C. Destri, H.J. de Vega, {\sl Non linear integral equation and
excited--states scaling functions in  the sine-Gordon model},
Nucl. Phys. {\bf B504} (1997) 621 and hep-th/9701107;


\bibitem{hub}
J. Hubbard, {\sl Electron correlation in narrow energy bands},
Proc. Roy. Soc. (London) {\bf A276} (1963) 238;

\bibitem{BEM}
E. Ercolessi, G. Morandi, A. M. Srivastava, A. P. Balachandran,
{\sl The Hubbard Model and Anyon Superconductivity}, World
Scientific;


\bibitem{KB}
A.Kl\"umper, R.Z. Bariev, {\sl Exact thermodynamics of the Hubbard
chain: free energy and correlation lengths}, Nucl. Phys. {\bf
B458} (1996) 623;

\bibitem{JKS}
G. J\"uttner, A. Kl\"umper, J.Suzuki, {\sl The Hubbard chain at
finite temperatures: ab initio calculations of Tomonaga-Luttinger
liquid properties}, Nucl. Phys. {\bf B522} (1998) 471 and
cond-mat/9711310;

\bibitem{DEGKKK}
T. Deguchi, F. H. L. Essler, F. G\"ohmann, A. Kl\"umper, V. E.
Korepin, K. Kusakabe, {\sl Thermodynamics and excitations of the
one-dimensional Hubbard model}, Phys. Rep. {\bf 331} (2000) 197
and cond-mat/9904398;

\bibitem{FR}
D. Fioravanti, M. Rossi, {\sl From finite geometry exact
quantities to (elliptic) scattering amplitudes for spin chains:
the 1/2-XYZ}, JHEP{\bf 08}(2005) 010 and hep-th/0504122;

\bibitem{TAK}
M. Takahashi, {\sl Thermodynamics of one-dimensional solvable
models}, Cambridge University Press;

\bibitem{woynar}
F. Woynarovich, H.P. Eckle, {\sl Finite-size corrections for the
low lying states of a half-filled Hubbard chain}, J. Phys. {\bf A
20} (1987) L443;

\bibitem {NZZ}
S. Sch\"afer-Nameki, M. Zamaklar, K. Zarembo, {\sl How accurate is
the quantum string Bethe Ansatz}, hep-th/0610250;

\bibitem{A}
P.W. Anderson, {\sl New Approach to the Theory of Superexchange
Interactions}, Phys. Rev. {\bf 115} (1959) 2;

\bibitem{GR}
I.S. Gradshteyn, I.M. Ryzhik, {\sl Table of integrals, series and
products}, Academic Press;

\bibitem{EP}
E.N. Economou, P.N. Poulopoulos, {\sl Ground state energy of the
half-filled one-dimensional Hubbard model}, Phys. Rev. {\bf B20}
(1979) 4756;

\bibitem{MV}
W. Metzner, D. Vohllhardt, {\sl Ground state energy of the
$d=1,2,3$ dimensional Hubbard model in the weak-coupling limit},
Phys. Rev. {\bf B39} (1989) 4462;

\bibitem{BO}
M. Beccaria, C. Ortix, {\sl Strong coupling anomalous dimensions
of ${\cal N} = 4$ super Yang-Mills}, JHEP {\bf 0609} (2006) 016
and hep-th/0606138;

\bibitem{BO2}
M. Beccaria, C. Ortix, {\sl AdS/CFT duality at strong coupling},
hep-th/0610215.


\end{thebibliography}
\end{document}